\documentclass[11pt]{article}

\usepackage{xcolor,fullpage}
\usepackage{pslatex}
\usepackage{graphicx}
\usepackage{fancyhdr}
\usepackage{amsmath}
\usepackage{amssymb}
\usepackage{textcomp}
\usepackage[square,sort,comma,numbers]{natbib}
\usepackage{subcaption}	
\usepackage{hyperref}
\usepackage{setspace}
\usepackage{cancel}
\usepackage{mathtools}
\usepackage[font=footnotesize]{caption}
\usepackage{rotating}
\usepackage{adjustbox}
\usepackage{multirow}
\usepackage{tabularx}
\usepackage{lipsum}
\usepackage{booktabs}
\usepackage{subfiles}
\usepackage{titling}
\usepackage{placeins}
\usepackage{authblk}
\usepackage{comment}
\usepackage{wrapfig}
\usepackage{chngcntr}

\newcolumntype{L}[1]{>{\raggedright\let\newline\\\arraybackslash\hspace{0pt}}m{#1}}
\newcolumntype{C}[1]{>{\centering\let\newline\\\arraybackslash\hspace{0pt}}m{#1}}
\newcolumntype{R}[1]{>{\raggedleft\let\newline\\\arraybackslash\hspace{0pt}}m{#1}}

\title{A deep learning approach to wall-shear stress quantification: From numerical training to zero-shot experimental application}
\author[1]{Esther Lagemann$^*$}
\author[2]{Julia Roeb}
\author[1]{Steven L. Brunton}
\author[1]{Christian Lagemann}
\affil[1]{\small AI Institute in Dynamic Systems, Department of Mechanical Engineering, University of Washington, Seattle, WA 98195, United States}
\affil[2]{\small Engler-Bunte-Institut, Simulation of Reacting Thermo-Fluid Systems, Karlsruhe Institute for Technology, Engler-Bunte-Ring 7, 76131 Karlsruhe, Germany}
\affil[$*$]{{\footnotesize corresponding author}}

\begin{document}
\date{}
\maketitle

\vspace{-2cm}

\begin{abstract}
\normalsize
The accurate quantification of wall-shear stress dynamics is of substantial importance for various applications in fundamental and applied research, spanning areas from human health to aircraft design and optimization. Despite significant progress in experimental measurement techniques and post-processing algorithms, temporally resolved wall-shear stress dynamics with adequate spatial resolution and within a suitable spatial domain remain an elusive goal. Furthermore, there is a systematic lack of universal models that can accurately replicate the instantaneous wall-shear stress dynamics in numerical simulations of multi-scale systems where direct numerical simulations are prohibitively expensive. To address these gaps, we introduce a deep learning architecture that ingests wall-parallel velocity fields from the logarithmic layer of turbulent wall-bounded flows and outputs the corresponding 2D wall-shear stress fields with identical spatial resolution and domain size. From a physical perspective, our framework acts as a surrogate model encapsulating the various mechanisms through which highly energetic outer-layer flow structures influence the governing wall-shear stress dynamics. The network is trained in a supervised fashion on a unified dataset comprising direct numerical simulations of statistically 1D turbulent channel and spatially developing turbulent boundary layer flows at friction Reynolds numbers ranging from $390$ to $1,500$. We demonstrate a zero-shot applicability to experimental velocity fields obtained from Particle-Image Velocimetry measurements and verify the physical accuracy of the wall-shear stress estimates with synchronized wall-shear stress measurements using the Micro-Pillar Shear-Stress Sensor for Reynolds numbers up to $2,000$. In summary, the presented framework lays the groundwork for extracting inaccessible experimental wall-shear stress information from readily available velocity measurements and thus, facilitates advancements in a variety of experimental applications.
\end{abstract}

\section{Introduction}
Turbulent wall-bounded fluid flows are of significant importance for numerous engineering~\cite{marusic2010predictive,Ricco2021,lagemann2023impact,mateling2023spanwise,lagemann2024extending} and biomedical applications~\cite{bellien2021,zhou2017,adamo2009biomechanical, tzima2005mechanosensory}, e.g., in the context of reducing the CO$_2$ emissions in the transportation sector or enhancing disease prevention and monitoring in human medicine. However, due to their high-dimensional, non-linear, and unsteady dynamics, we still lack a comprehensive understanding of these flows. One particular quantity of interest is the wall-shear stress since it is a measure of the friction forces and the flow-induced dynamic loads acting on the surface. However, measuring the wall-shear stress in experimental settings is still a significant - often even an unfeasible - challenge~\cite{orlu2020instantaneous,lagemann2024}. Therefore, the present work introduces a modern deep learning based algorithm to predict the wall-shear stress distribution based on accessible velocity measurements, which are usually obtained from Particle-Image Velocimetry (PIV) experiments~\cite{raffel_particle_2018,scharnowski2020}. Recent advances in PIV image processing~\cite{lagemann2021,lagemann2022,lagemann2024challenges} have demonstrated how a deep optical flow network can be used to derive highly accurate wall-shear stress dynamics from PIV measurements in the wall-normal plane in which the viscous sublayer is well resolved~\cite{lagemann2024}. However, such an approach requires proper experimental conditions with a sufficiently high spatial resolution, which can be challenging especially for high Reynolds number flows and in the presence of experimental burdens like optical access or reflections at the wall. Therefore, a well-designed model capable of predicting the time-dependent dynamics of the spatial wall-shear stress distribution solely based on readily available wall-parallel measurement data at a certain distance from the wall is of tremendous value. The foundation for the success of such a model is rooted in a vast literature~\cite{agostini2014,baars2015,marusic2007,marusic2017,mathis2009,pathikonda2019,maeteling2020,maeteling2022} showing that the dominant dynamics of the inner layer of a turbulent wall-bounded flow and consequently, of the wall-shear stress, are imposed by highly energetic outer-layer flow structures located in the logarithmic region. Thus, based on this inner-outer interaction, it can be hypothesized that outer-layer velocity fields contain sufficient information to derive the governing wall-shear stress dynamics.
Moreover, recent advances in machine learning are revolutionizing how we approach the challenging analysis of complex fluid dynamical systems characterized by high-dimensionality, non-linearity, and multi-scale features~\cite{lagemann2023invariance,erichson2020,brunton2020,guemes2021,vinuesa2022,vinuesa2023,raissi2019,eivazi2024physics}, establishing a promising foundation for their applicability in the present work. Precisely, we developed a deep neural operator network specifically designed to learn a mapping function from 2D wall-parallel velocity fields located in the outer layer of turbulent wall-bounded flows to the instantaneous and spatially resolved wall-shear stress distribution. Trained on direct numerical simulation (DNS) data of turbulent channel and turbulent boundary layer flows at friction Reynolds numbers ranging from $Re_\tau \approx 390$ to $1,500$, the generalization of our framework is evidenced with an experimental setup using simultaneous PIV measurements in the outer layer and wall-shear stress measurements with the Micro-Pillar Shear-Stress Sensor (MPS$^3$)~\cite{grosse2008b,grosse2009,nottebrock2012,geurts2014,liu2019,maeteling2020exp} at friction Reynolds numbers up to $Re_\tau \approx 2,000$. 

The idea of using a neural architecture to predict the wall-shear stress was explored in a few former studies~\cite{su2020,gharleghi2022,suk2021,balasubramanian2023}. In a medical context, the first attempt to predict the wall-shear stress in a stenosed coronary artery using machine learning was proposed by Su et al.~\cite{su2020}. Based on the geometry of synthetic coronary artery models, a convolutional autoencoder estimates the wall-shear stress distribution. A subsequent study extended this framework to 3D geometries~\cite{suk2021}. Going beyond purely geometric input data, Gharleghi et al.~\cite{gharleghi2022} invoke a steady state solution of the cardiac flow field to predict the wall-shear stress in bifurcations using convolutional neural networks. However, approaches solely based on geometric features and a time-averaged flow field are unlikely to succeed in predicting instantaneous wall-shear stress information as required for more detailed hemodynamics analyses. Therefore, an emerging research direction in medical science focuses on deriving wall-shear stress data from 4D flow cardiac magnetic resonance imaging with machine learning~\cite{ferdian2022wssnet,garrido2022machine}. However, the coarse spatial resolution of state-of-the-art cardiac magnetic resonance imaging data is a significant burden for deriving highly resolved wall-shear stress dynamics~\cite{arzani2021data,shone2023deep}. \\

\begin{figure}[t!]
\centering
\includegraphics[width=0.99\textwidth]{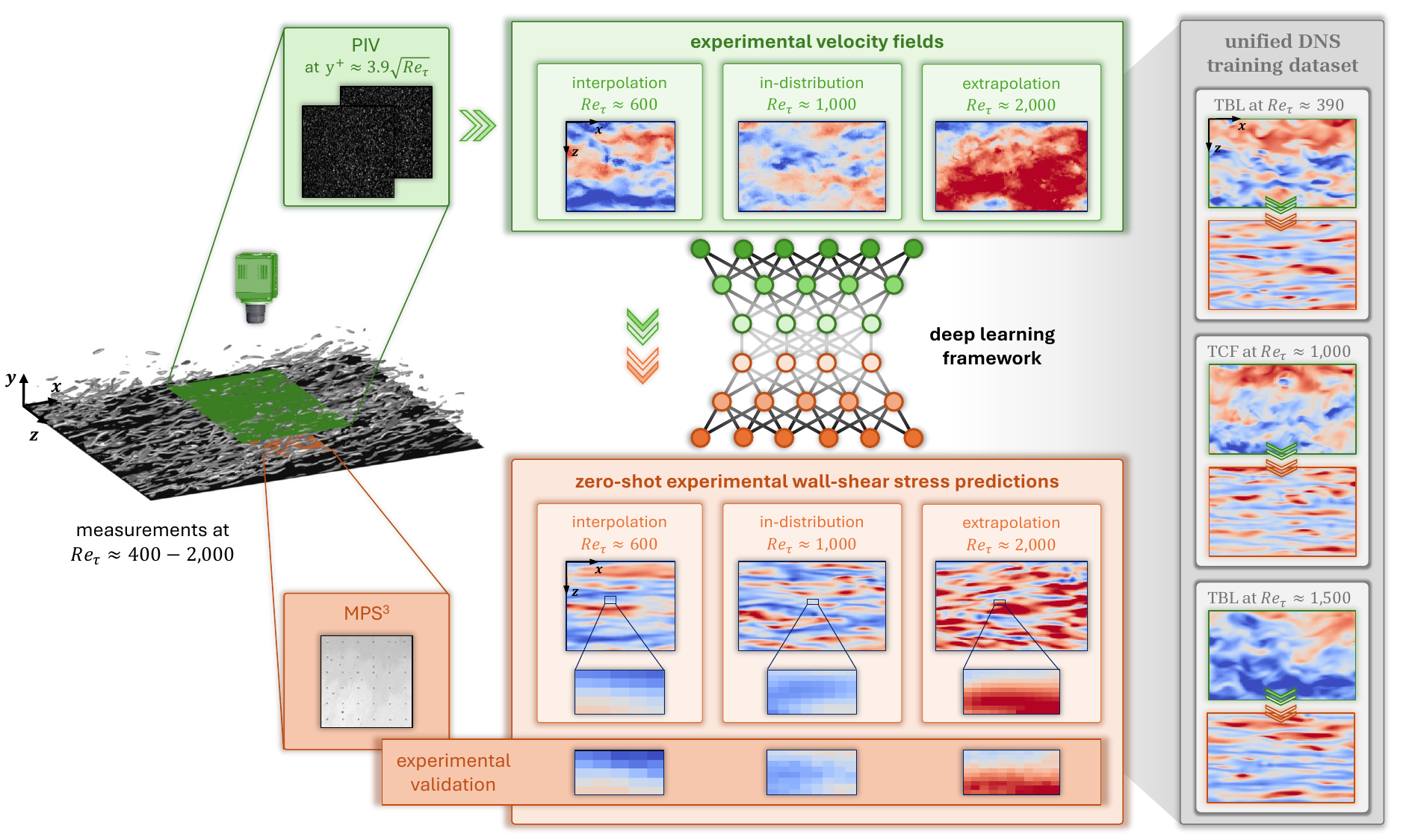}
  \caption{\textbf{Workflow using the proposed deep learning framework to derive instantaneous 2D wall-shear stress dynamics from outer-layer velocity fields.} The left-hand side shows a sketch of the experimental setup combining Particle-Image Velocimetry (PIV) based velocity measurements at a wall-normal distance of $y^+ \approx 3.9 \sqrt{Re_\tau}$ with wall-shear stress measurements using the Micro-Pillar Shear-Stress Sensor (MPS$^3$). The measurements are conducted in fully developed turbulent channel flows at friction Reynolds numbers ranging from $Re_\tau \approx 400$ to $2,000$. The experimental velocity fields are processed by the deep neural network (displayed in a simplified sketch) and the physical accuracy of the predicted wall-shear stress fields is verified with the measured wall-shear stress distributions. Our framework shows a strong generalization ability with respect to in-distribution data, i.e., Reynolds numbers that were included in the training dataset (e.g., $Re_\tau \approx 1,000$), interpolated data, i.e., Reynolds numbers within the range of the training distribution ($Re_\tau \approx 600$), as well as extrapolated data, i.e., Reynolds numbers outside the training distribution ($Re_\tau \approx 2,000$). Moreover, it is important to note that the utilization of experimental data is a zero-shot application meaning that the neural network is never trained on the specific characteristics of experimental measurement data. Precisely, the network is solely trained on Direct Numerical Simulation (DNS) data of turbulent channel (TCF) and turbulent boundary layer (TBL) flows at three Reynolds numbers ($Re_\tau \approx 390, 1,000$ and $1,500$). Examples of the respective instantaneous velocity and wall-shear stress data are provided on the right-hand side in the gray box.}
  \label{fig:motivation}
\end{figure}

Most recently, attempts to predict high-resolution instantaneous wall-shear stress fields have been made in the fluid dynamics community. Balasubramanian et al.~\cite{balasubramanian2023} demonstrated a successful estimation of instantaneous wall quantities, i.e., wall-shear stress and wall pressure, from velocity information at several distances to the wall based on DNS data of open turbulent channel flow. The authors compare the performance of a fully convolutional network to a newly proposed R-Net at friction Reynolds numbers of $Re_\tau \approx 180$ and $550$. For velocity fields extracted from the buffer layer, the R-Net predictions are very accurate but an increasing deviation from the true DNS data was observed for larger wall distances. Building upon these initiatives, the present work incorporates three novel aspects related to the Reynolds number regime, the generalization ability across flow problems and Reynolds numbers, as well as an experimental validation, which are crucial for the advancement of deep learning driven wall-shear stress quantification for real-world applications. First, we focus on a regime with Reynolds numbers of up to $Re_\tau \approx 2,000$ in which the large-scale flow structures from the outer layer interact with the flow features close to the wall. This established inner-outer interaction has a substantial impact on the wall-shear stress dynamics~\cite{marusic2007,ganapathisubramani2012,maeteling2022} and becomes increasingly important with higher Reynolds numbers~\cite{marusic2010_peak}. Thus, by inherently learning these multi-scale interaction processes, our architecture provides a valuable foundation for prospective extensions to even higher Reynolds number flows relevant for, e.g., drag reduction research for prospective aircraft applications. Second, we demonstrate a zero-shot applicability to experimental data using a network that is entirely trained on numerical data. The ability to validate that the framework can effectively manage the inevitable measurement uncertainty intrinsic to experimental data represents a substantial transition from academic research to practical applications. It builds the foundation for enhancing experimental measurements with pre-trained machine learning models as well as for developing and verifying novel models for numerical simulations, e.g., in terms of wall-modeled large-eddy simulations. Third, our deep neural network is the first framework for wall-shear stress estimation that reliably generalizes across flow conditions. That is, we demonstrate a successful performance for a training dataset comprising statistically 1D turbulent channel as well as spatially developing turbulent boundary layer flows for friction Reynolds numbers in the range of $390 \le Re_\tau \le 1,500$. Moreover, we verify the physical significance of the predicted wall-shear stress distributions for experimental data with respect to in-distribution and intermediate Reynolds numbers as well as an extrapolation to $Re_\tau \approx 2,000$. Such a generalization ability allows straightforward applications in a variety of experimental studies and thus, the proposed neural architecture constitutes a great asset for the fluid dynamics community. \\
To summarize, the novelty of our contribution is based on the following key aspects:
\begin{itemize}
    \item The neural modeling of the inner-outer interaction mechanisms allows accurate wall-shear stress predictions from outer-layer velocity fields.
    \item The validation of a zero-shot applicability to experimental data demonstrates how the combination of our deep learning framework with PIV measurements provides accurate -~and otherwise inaccessible~- instantaneous experimental wall-shear stress distributions within a large spatial domain. 
    \item The generalization ability across Reynolds numbers and flow configurations allows an easy and straightforward application to a variety of experimental settings.     
\end{itemize}

The paper is structured as follows. First, the numerical and experimental datasets are described in section~\ref{sec:datasets}. Subsequently, the proposed neural architecture is introduced in section~\ref{sec:architecture}. The results (section~\ref{sec:results}) are divided into two main sections. Initially, the proposed neural network is solely trained on numerical data of a turbulent channel flow at a single Reynolds number of $Re_\tau \approx 1,000$ (section~\ref{sec:results1}). This provides an optimal training environment since only the particular properties of a single flow condition have to be learned. We analyze the results of this framework in great detail by studying the in-distribution results with respect to numerical test data (section~\ref{subsec:results1_DNS}) and the zero-shot applicability to experimental data capturing the same flow conditions (section~\ref{subsec:results1_exp}). In the second part (section~\ref{sec:results2}), the proposed architecture is trained on a multi-configuration dataset comprising a turbulent channel flow at $Re_\tau \approx 1,000$ and two turbulent boundary layer flows at $Re_\tau \approx 390$ and $Re_\tau \approx 1,500$. The performance is evaluated using in-distribution numerical test data of all three configurations as well as experimental data at similar Reynolds numbers. Furthermore, we study how well the network is able to predict wall-shear stress fields for an intermediate ($Re_\tau \approx 600$) and a higher Reynolds number ($Re_\tau \approx 2,000$) using experimental data. A conclusive discussion is given in section~\ref{sec:discussion}. 

\section{Numerical and experimental datasets} \label{sec:datasets}
Three direct numerical simulation (DNS) datasets are used in the present study, namely a fully developed turbulent channel flow (TCF) at a friction Reynolds number of $Re_\tau = u_\tau h/\nu \approx 1,000$ (section~\ref{subsec:DNS_TCF}) and turbulent boundary layer (TBL) flows at $Re_\tau = u_\tau \delta/\nu \approx 390$ and $Re_\tau \approx 1,500$ (section~\ref{subsec:DNS_TBL}), where $u_\tau$ denotes the friction velocity, $h$ the channel half height, $\delta$ the boundary layer thickness, and $\nu$ the kinematic viscosity. The TCF is provided by the Johns Hopkins Turbulence Databases~\cite{graham2016,li2008,perlman2007data} and the TBL flows are identical to the reference configurations used in \cite{mateling2023spanwise,lagemann2023impact}. Moreover, experimental TCF measurements at five Reynolds numbers have been conducted (section~\ref{subsec:exp_TCF}). They cover the three Reynolds numbers of the DNS datasets, one intermediate flow condition at $Re_\tau \approx 600$ and one higher Reynolds number flow at $Re_\tau \approx 2,000$. A thorough comparison between the numerical and the experimental TCF data at $Re_\tau \approx 1,000$ is provided in section~\ref{subsec:Datasets_compare}. \\
For all datasets, wall-parallel streamwise velocity fields are extracted at the outer-layer wall-normal location where the large-scale motions are most energetic ($y^+ \approx 3.9 \sqrt{Re_\tau}$~\cite{marusic2010_peak}). Thus, these flow structures are expected to have the most significant impact on the wall-shear stress dynamics. For the numerical datasets, corresponding wall-shear stress fields are extracted at the same spatial location and with the same spatial resolution. Experimental wall-shear stress information is available in a limited spatial extent to validate the physical correctness of the neural wall-shear stress predictions. The streamwise velocity and wall-shear stress fields used for training are zero-mean quantities in inner units. 

\subsection{Numerical turbulent channel flow} \label{subsec:DNS_TCF}
DNS data of a fully developed TCF at a friction Reynolds number of $Re_\tau \approx 1,000$ is provided by the Johns Hopkins Turbulence Databases~\cite{graham2016,li2008,perlman2007data} and the interested reader is referred to these references for further information on the numerical simulations. For the present study, $800$ time steps with a temporal separation of $\Delta t^+ = 1.625$ are used. Each snapshot contains $x \times z = 2,048 \times 1,536$ cells, where $x$ denotes the streamwise and $z$ the spanwise direction. It is subdivided into $12$ fields measuring $x \times z = 128 \times 128$. In each spatial direction, these $12$ fields are separated by $384$ cells to increase the variability within the samples. Thus, the total number of available samples is $9,600$ with a sample size of $x \times z = 128 \times 128$ and a physical resolution of $\Delta x^+ = 12.26$ and $\Delta z^+ = 6.13$. The extracted velocity fields are located at $y^+ \approx 3.9 \sqrt{Re_\tau} \approx 123.4$.

\subsection{Numerical turbulent boundary layer flows} \label{subsec:DNS_TBL}
The TBL flows are identical to the reference configurations used in \cite{mateling2023spanwise,lagemann2023impact} and additional information regarding the computational setup and the numerical solver are given in \cite{Albers2020FTaC,fernex2020,mateling2023spanwise,lagemann2023impact}. The velocity fields are extracted at $y^+ \approx 78.8$ for $Re_\tau \approx 390$ and at $y^+ \approx 155.6$ for $Re_\tau \approx 1,500$. The datasets have an initial resolution of $\Delta t^+ = 3.49, \Delta x^+= 9.94, \Delta z^+ = 4.52$ ($Re_\tau \approx 390$) and $\Delta t^+ = 3.21, \Delta x^+= 11.58, \Delta z^+ = 7.72$ ($Re_\tau \approx 1,500$). Both datasets are interpolated onto the same grid as the numerical TCF using cubic spline functions since the convolutional neural architecture requires a consistent spatial resolution when trained on all three numerical datasets. After interpolation, velocity and wall-shear stress fields of size $x \times z = 128 \times 128$ are extracted to build the final datasets. In total, $5,160$ samples are extracted for each configuration.

\subsection{Experimental turbulent channel flows} \label{subsec:exp_TCF}
\begin{figure}[t!]
\centering
\includegraphics[width=0.8\textwidth]{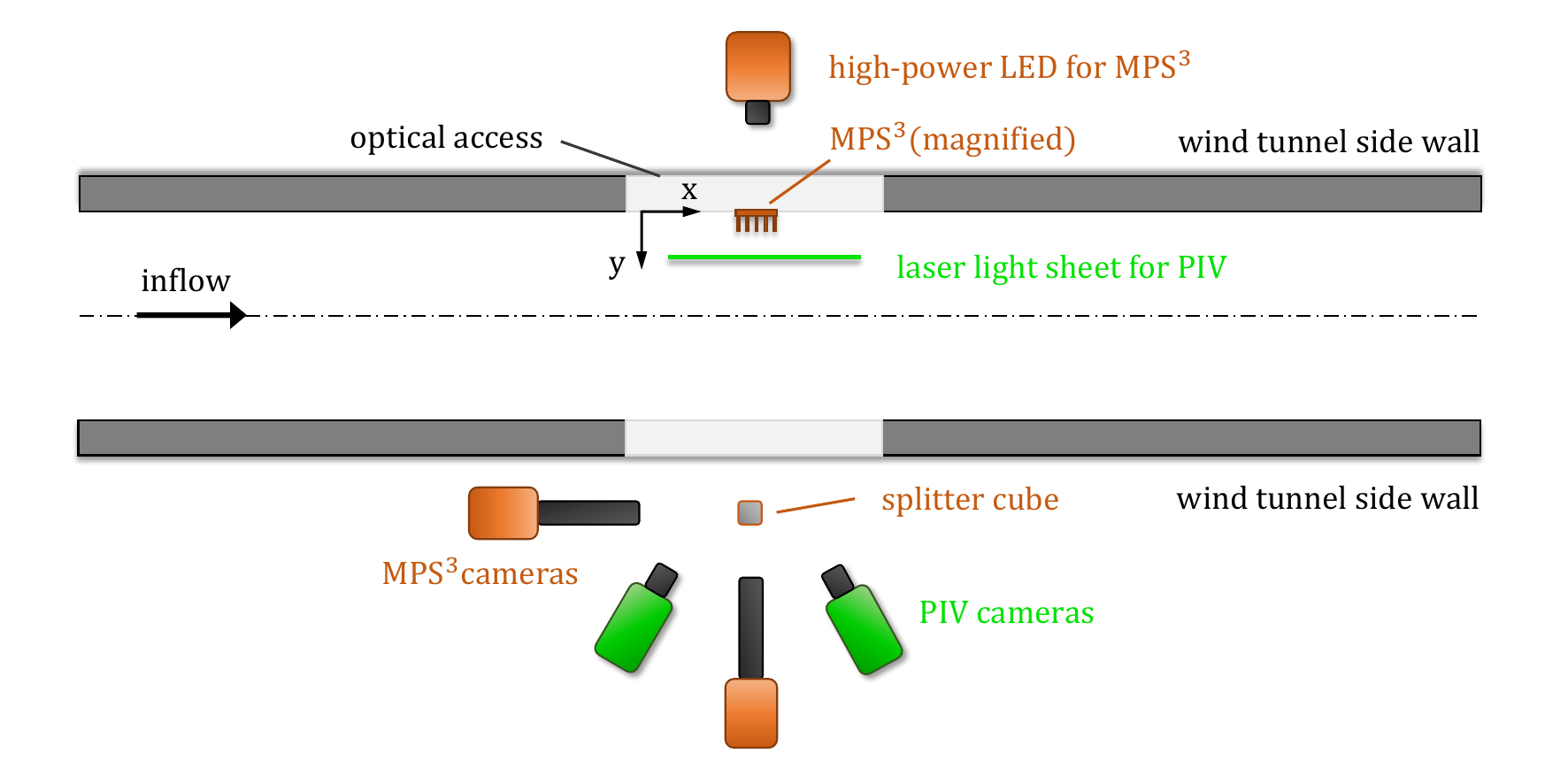}
  \caption{\textbf{Sketch of the experimental setup comprising simultaneous and synchronized stereo PIV and MPS$^3$ measurements in fully developed TCFs.} The PIV measurement plane is located at $y^+ \approx 3.9 \sqrt{Re_\tau}$ parallel to the channel wall and illuminated by a pulsed laser light sheet. Two cameras with Scheimpflug adapters are focused onto the measurement plane. Simultaneously, two cameras equipped with microscope lenses observe the MPS$^3$ in a top view using a shared beam splitter. The MPS$^3$ is illuminated in a backlight setup using an LED with a distinctly different wavelength than the laser light. Thus, by using bandpass filters for all cameras, both measurement systems are optically separated.}
  \label{fig:exp_setup}
\end{figure}

Simultaneous and synchronized experimental velocity and wall-shear stress measurements were conducted in a fully developed TCF with an aspect ratio of $20$ at five friction Reynolds numbers between $Re_\tau \approx 400$ and $2,025$. The experimental datasets are used to validate the performance of the deep learning framework with respect to measurement data. Thus, the measurements were targeted to capture the same flow conditions as the numerical datasets, however, small deviations in the experimental settings are unavoidable. Furthermore, two additional configurations are considered to test the performance with an intermediate ($Re_\tau \approx 610$) and an extrapolated ($Re_\tau \approx 2,025$) dataset. All relevant experimental measures are summarized in table~\ref{tab:expMeasures}. \\
The experimental setup is sketched in figure~\ref{fig:exp_setup} and is similar to the setup introduced in \cite{maeteling2020exp}. Stereoscopic PIV measurements in a wall-parallel plane at $y^+ \approx 3.9 \sqrt{Re_\tau}$ are synchronized with wall-shear stress measurements using the Micro-Pillar Shear-Stress Sensor (MPS$^3$)~\cite{grosse2008b,grosse2009,nottebrock2012,geurts2014,liu2019,maeteling2020exp}. The sensor consists of flexible, cylindrical structures ("micro pillars"), which are immersed in the viscous sublayer and bend due to the experienced fluid forces. A preceding calibration relates the micro-pillar deflection to the instantaneous wall-shear stress. Further details on the measurement principle and the data post-processing are described in \cite{maeteling2020exp}. 

\begin{table}[b!]
\centering
\renewcommand{\arraystretch}{1.3} 
\newcolumntype{Y}{>{\centering\arraybackslash}X} 
\begin{tabularx}{\textwidth}{p{0.9cm} p{2.4cm}| Y Y Y Y Y }
\toprule
& & $Re_\tau \approx 400$ & $Re_\tau \approx 610$ & $Re_\tau \approx 970$ & $Re_\tau \approx 1,470$ & $Re_\tau \approx 2,025$ \\
\midrule
\multirow{5}{*}{\centering PIV} & $y^+$ & 79.32 & 98.00  & 122.09 & 147.25 & 174.15 \\
& $\Delta x^+, \Delta z^+$ & 2.12, 1.87 & 3.27, 2.88 & 5.17, 4.56 & 7.85, 6.93 & 10.80, 9.53 \\
& $\Delta x^+_{int}, \Delta z^+_{int}$ & 12.26, 6.13 & 12.26, 6.13 & 12.26, 6.13 & 12.26, 6.13 & 12.26, 6.13 \\
& $x^+_{int} \times z^+_{int}$ & 515 $\times$ 442 & 797 $\times$ 681 & 1263 $\times$ 1116 & 1913 $\times$ 1643 & 2637 $\times$ 2263 \\ \midrule
\multirow{6}{*}{\centering MPS$^3$} & $x \times z$ & 16 $\times$ 9 & 16 $\times$ 9 & 16 $\times$ 9 & 17 $\times$ 8 & 17 $\times$ 8 \\
& $x_{int} \times z_{int}$ & 4 $\times$ 3 & 5 $\times$ 5 & 8 $\times$ 8 & 8 $\times$ 7 & 11 $\times$ 10 \\
& $\Delta x^+, \Delta z^+$ & 2.38, 2.38 & 3.68, 3.68 & 5.81, 5.81 & 5.89, 5.89 & 8.10, 8.10 \\
& $\Delta x^+_{int}, \Delta z^+_{int}$ & 12.26, 6.13 & 12.26, 6.13 & 12.26, 6.13 & 12.26, 6.13 & 12.26, 6.13 \\
& $x^+_{int} \times z^+_{int}$ & 25 $\times$ 18  & 49 $\times$ 25 & 86 $\times$ 43 & 86 $\times$ 37 & 123 $\times$ 55 \\
\bottomrule
\end{tabularx}
\caption{\textbf{Experimental details of the synchronized PIV and MPS$^3$ measurements.} The given measures are the inner-scaled wall-normal distance of the PIV measurements ($y^+$), the initial data resolution in streamwise and spanwise direction prior to interpolation onto the DNS grid ($\Delta x^+, \Delta z^+$), the original number of wall-shear stress vectors prior to interpolation ($x \times z$), and the data size in inner units ($x^+ \times z^+$). The subscript $(\cdot)_{int}$ indicates the interpolated quantities used during inference with the deep learning framework.}
\label{tab:expMeasures}
\end{table}

The PIV setup consists of two \textit{Photron FASTCAM SA3} high-speed cameras equipped with Scheimpflug adapters and $100$~$mm$  F/$2.0$ \textit{Zeiss} macro lenses. The cameras are synchronized to a \textit{Quantronix Darwin Duo 100} high-speed laser operated at a frame rate of $2,000$~Hz. Two different MPS$^3$ sensors are used to ensure that the micro pillars are embedded within the viscous sublayer and possess adequate sensitivity to the flow conditions for all configurations. Thus, for $Re_\tau \approx 400, 610, 970$, the MPS$^3$ consists of $x \times z = 16 \times 9$ micro pillars with a height of $L_p = 300$~$\mu m$ and a diameter of $D_p = 22$~$\mu m$ equidistantly spaced by $L_p$ in streamwise and spanwise direction. For $Re_\tau \approx 1,470$ and $2,025$, the micro pillar length is reduced to $L_p = 200$~$\mu m$ with a diameter of $D_p = 14$~$\mu m$ and $x \times z = 17 \times 8$ micro pillars. The pillar tip deflection is observed in a top view by two \textit{pco.dimax HS4} high-speed cameras equipped with \textit{K2/SC long-distance microscope} lenses. Due to spatial constraints, a splitter cube is used for optical access. The MPS$^3$ is illuminated by a pulsed high-power LED system \textit{LPS V3} in a backlight configuration at a frame rate of $1,000$~Hz. Since PIV processing requires an image pair but the MPS$^3$ relies on only a single image, which is compared to the reference image at zero flow, its frame rate is halved. To optically separate both measurement systems, the laser and the LED are operated at different wavelengths and the cameras are equipped with corresponding bandpass filters. In total, $1,000$ data samples are recorded for each configuration. 

The PIV data are evaluated with \textit{PascalPIV}~\cite{marquardt2019, marquardt2020experimental} using a multi-grid approach with a final interrogation window size of $16 \times 16$~px$^2$ and an overlap of $75$~\%. This results in a final velocity vector spacing of $4 \times 4$~px$^2$, the physical resolution of which is given in table~\ref{tab:expMeasures} for all five Reynolds numbers. For the wall-shear stress measurements, a post-processing as described in \cite{maeteling2020exp} is applied and the resulting data sizes and resolutions are provided in table~\ref{tab:expMeasures}. In a final step, all datasets are interpolated onto the same grid as the numerical TCF data ($\Delta x^+ = 12.26$, $\Delta z^+ = 6.13$) using cubic splines since a consistent physical resolution is required for inference. The interpolation of the wall-shear stress values results in distributions of $x \times z = 4 \times 3$ values for the smallest Reynolds number and $x \times z = 11 \times 10$ for the highest Reynolds number. Although these fields do not cover the entire spatial domain, their spatial information is sufficient to verify the physical significance of the neural wall-shear stress predictions.   

\subsection{Comparison between numerical and experimental datasets}\label{subsec:Datasets_compare}
Although the experiments were targeted to resemble the flow conditions of the numerical data, small inaccuracies and uncertainties are unavoidable in experimental settings. Therefore, we briefly highlight the most significant differences between both datasets for the TCFs at $Re_\tau \approx 1,000$. This provides a clear foundation for assessing the performance of the neural architecture with respect to the zero-shot applicability to experimental data. \\
Since both datasets capture a fully developed TCF, any influence of the inflow conditions is negligible. The experimental facility features an aspect ratio of $20$ such that the channel's side walls do not influence the flow along the centerline where the measurements were conducted. Periodic boundary conditions in the DNS exclude side wall effects, too. The friction Reynolds number of the DNS is $Re_\tau \approx 1,000$, whereas the experimental dataset captures the flow field at $Re_\tau \approx 970$. Moreover, a small offset of the wall-normal position of the wall-parallel velocity fields exists ($y^+ = 123.38$ in the DNS and $y^+ = 122.09$ in the experiments). However, these differences do not impact the observed flow field characteristics. To investigate the variation within the outer-layer velocity fields in a statistical sense, we apply the 2D Noise-Assisted Multivariate Empirical Mode Decomposition (2D NA-MEMD)~\cite{maeteling2022} and subsequently perform a Fast Fourier Transform (FFT) on each mode to inspect the power spectra. In essence, the 2D NA-MEMD decomposes the velocity fields into scale-based modal representations, which are called Intrinsic Mode Functions (IMFs), using a data-driven iterative procedure. By simultaneously decomposing the numerical and the experimental data, we ensure that the resulting IMFs capture the same range of scales and reveal similarities as well as differences between numerical and experimental distributions. Thus, we can analyze which range of scales possesses the largest difference, e.g., due to measurement noise or bias. More details regarding the 2D NA-MEMD and its application to fluid flow data can be found in \cite{maeteling2022,mateling2023spanwise,lagemann2023impact,lagemann2024}. 

\begin{figure}[t]
\centering
\includegraphics[width=0.99\textwidth]{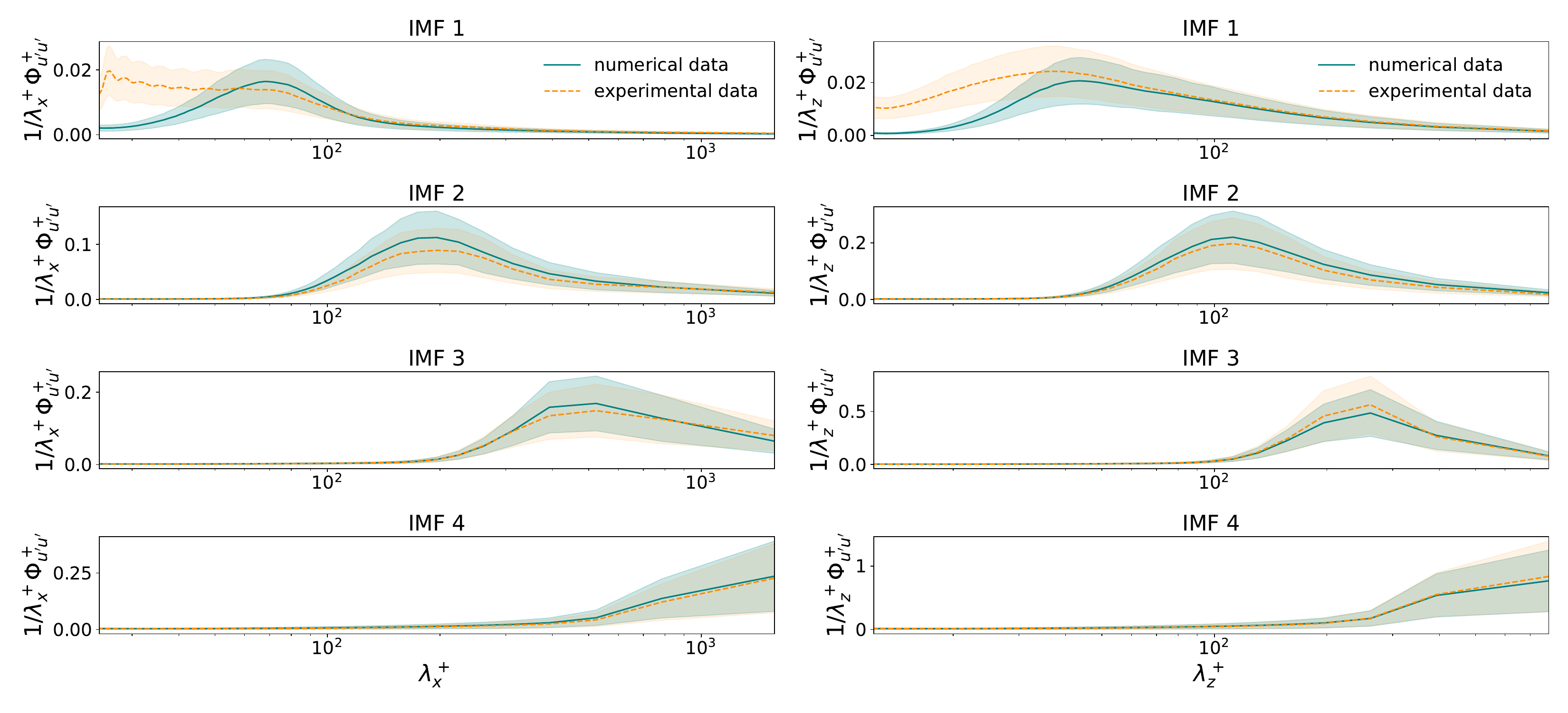}
  \caption{\textbf{Pre-multiplied power spectra of the streamwise velocity fluctuations of the TCF at $\mathbf{Re_\tau \approx 1,000}$ decomposed into modal representations using the 2D NA-MEMD.} The 2D NA-MEMD is applied to the streamwise velocity fluctuations of the numerical and the experimental datasets to perform a simultaneous, scale-based decomposition into modal representations (IMFs). Sorted with respect to the inherent scale sizes, the IMFs reveal which range of scales are common between both datasets and how their energetic content behaves. By subsequently applying an FFT to all IMFs and taking the mean (lines) and the standard deviation (shaded regions) across all samples, the inner-scaled pre-multiplied power spectra given in this figure are obtained. The left column provides spectra as a function of the streamwise wavelengths $\lambda_x^+$ and the right column as a function of the spanwise wavelengths $\lambda_z^+$. The most distinct difference between experimental and numerical spectra occurs in IMF$1$, in which very small wavelengths of the experimental data comprise a significant amount of energy. This non-physical energy redistribution arises due to measurement noise and also slightly impacts the statistics of IMF$2$.}
  \label{fig:OLdistribution}
\end{figure}

Figure~\ref{fig:OLdistribution} shows the respective inner-scaled pre-multiplied power spectra of the IMFs, which are averaged across the entire datasets, as a function of the streamwise $\lambda_x^+$ and the spanwise $\lambda_z^+$ wavelengths. The corresponding standard deviations are indicated by the shaded regions. This representation reveals a significant amount of small-scale measurement noise inherent to the experimental data, which manifests in the first IMF. With increasing scale size, i.e., higher mode number, the difference between numerical and experimental datasets decreases. Thus, the most severe variation is rooted in the smallest scales. Although the spatial resolution of the experimental dataset is higher than the DNS resolution, it does not necessarily mean that the small scales are well resolved. The resolution of the physical features depends on the chosen pulse distance of the PIV measurements, which determines how far the particles have traveled between the first and the second exposure. If the measurements target the governing large-scale flow dynamics, which was the focus of this setup, this comes at the cost of loosing small-scale information.

\section{Deep learning architecture} \label{sec:architecture}
\begin{figure}[h!]
\centering
\includegraphics[width=0.7\textwidth]{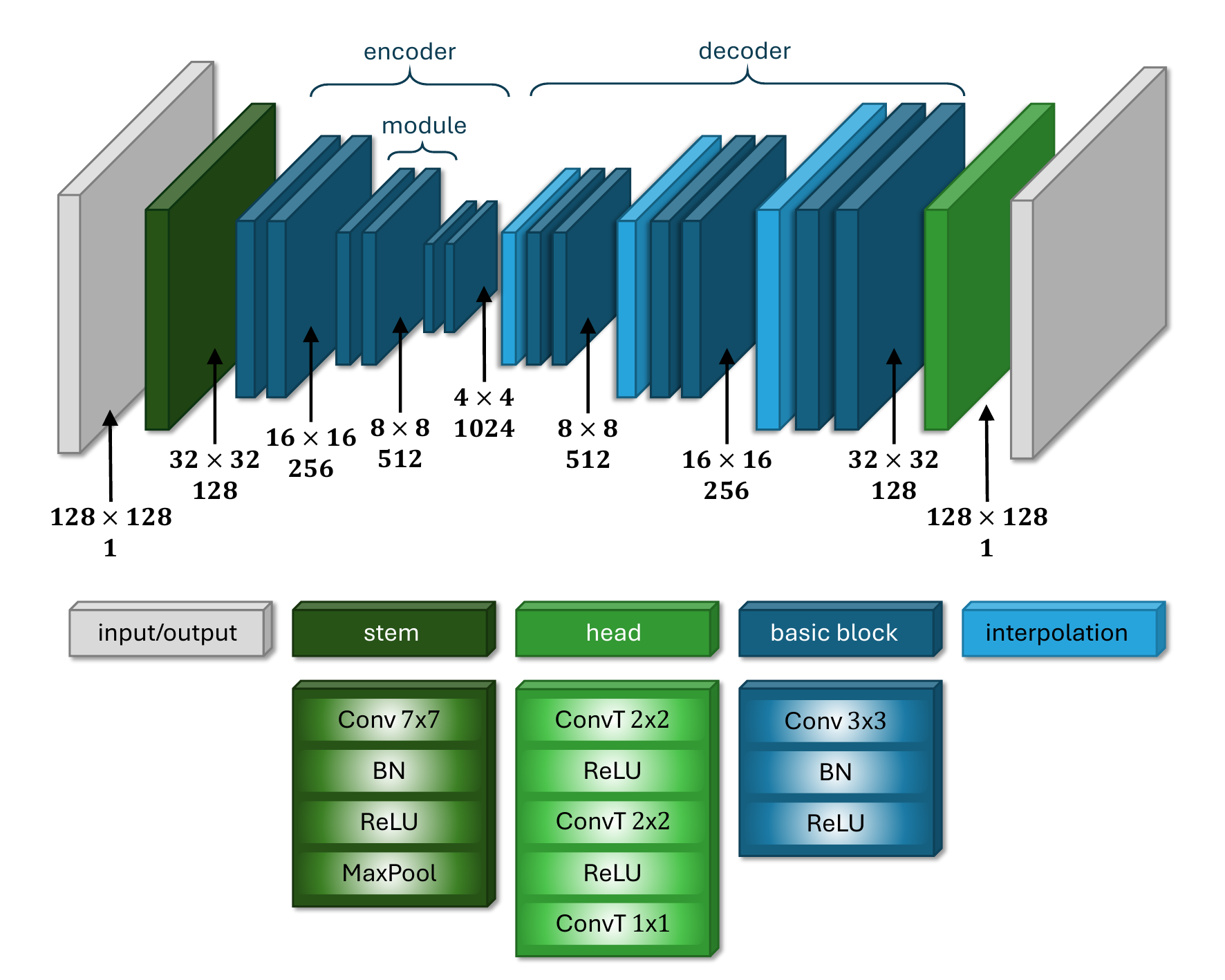}
  \caption{\textbf{Sketch of the proposed deep learning architecture.} A streamwise velocity fluctuation field is evaluated by the autoencoder, which predicts the corresponding wall-shear stress field at the same image resolution. The velocity field is first encoded by a stem block, followed by three basic modules. Each module consists of several basic blocks and the ReLU activation function is omitted at the last stage of each module. During decoding, each basic module starts with an interpolation layer to double the image dimension. The three decoder basic modules are followed by a head block, which outputs the wall-shear stress prediction. The network is trained in a supervised fashion using a loss function based on the L$2$ norm between ground truth and prediction. \\\hspace{\textwidth} The output dimension of each module is indicated using the following notation: spanwise $\times$ streamwise image dimension in the upper row, feature space dimension in the lower row. Abbreviations: Convolution (Conv), batch normalization (BN), rectified linear unit (ReLU), max pooling (MaxPool), transposed convolution (ConvT). The kernel size of the convolutions is provided after the convolution operator.}
  \label{fig:architecture}
\end{figure}

The proposed framework is a neural operator network based on a convolutional autoencoder architecture, which is sketched in figure~\ref{fig:architecture}. A wall-parallel streamwise velocity fluctuation field of dimensions $128 \times 128$~px$^2$ ($x^+ \times z^+ \approx 1,570 \times 785$) is evaluated the network, which outputs the corresponding wall-shear stress fluctuation field with the same dimension and spatial resolution. The input is first processed by a stem block, which includes a convolution with a kernel size of $7 \times 7$ and a stride of $2$, followed by a batch normalization, ReLU activation, and max pooling with a kernel size of $3$ and a stride of $2$. The output of the stem block is a spatially compressed matrix of $32 \times 32$~px$^2$ with $128$ feature maps. It is then processed by several basic blocks, which contain a convolution with a kernel size of $3 \times 3$, a batch normalization, and ReLU activation. The first basic block of each module (same-sized basic blocks in figure~\ref{fig:architecture} build a module) applies a convolution with a stride of $2$ and a doubling of the feature space, whereas the subsequent blocks have a stride of $1$ and preserve the feature space dimension. The ReLU activation is omitted at the output of each module. The compressed latent representation has a spatial dimension of $4 \times 4$~px$^2$ with $1024$ feature maps. Each module of the decoder contains a preceding interpolation layer, which upsamples the spatial dimension by a factor of $2$ using nearest neighbors interpolation. Our studies showed that this type of upsampling in the early stages of decoding improves the resolution of small-scale features compared to the often used transposed convolution. All basic blocks of the decoder have a stride of $1$ and the first basic block in each module halves the feature space. In the final stage, the head block performs three transposed convolutions with subsequent ReLU activation except for the final output. The network is trained in a supervised fashion using the L$2$ norm between the wall-shear stress prediction and the ground truth as the training loss. 

The exact number of basic blocks contained in each encoding and decoding module is the result of a hyperparameter study. The encoder performs best with four basic blocks in the first two basic modules and two basic blocks in the third module. For the decoding path, we employ four basic blocks in each network stage. Moreover, we studied the impact of using residual layers within each module (ResNet) as well as cross-connections between the encoder and decoder (Unet). By fine-tuning each of these networks with hyperparameter studies, we achieved a similar performance across all architectures. For the sake of readability, we therefore only discuss results related to the basic autoencoding architecture and provide further details and comparisons with respect to the other architectures in the \textit{Appendix}. 

The network training is performed on four \textit{NVIDIA A100 GPUs}. We used a batch size of $7$ and a dropout ratio of $0.3$ to prevent overfitting. An Adam optimizer with an initial learning rate of $9.7 \cdot 10^{-5}$ and weight decay of $1.7 \cdot 10^{-4}$ was combined with a scheduler that reduces the learning rate by a factor of $0.2$ when the validation loss is not improving after $10$ epochs. In total, training was performed for $500$ epochs with a minimum learning rate of $10^{-10}$. All training specific parameters have been optimized using extensive hyperparameter studies. \\
Two different training setups are used in the present study. First, we solely train the neural network on the TCF data at $Re_\tau \approx 1,000$ to investigate the capabilities of the network in the optimal setting (i.e. only a single flow condition). Second, we use all three configurations, i.e., the TCF and both TBL flows, to analyze the generalization ability of the network across flow conditions and slightly changing flow problems (statistically 1D vs. statistically developing flow in streamwise direction). For the first scenario, the numerical TCF dataset described in section~\ref{subsec:DNS_TCF} is randomly divided into a training dataset containing $80$~\% of the snapshots ($7,680$ samples), a validation dataset comprising $10$~\% of the snapshots ($960$ samples), and a test dataset with the remaining $10$~\% ($960$ samples). While a validation is performed at the end of each epoch, the test dataset is only used after the entire training procedure is terminated. Thus, it serves as a verification for the network performance on unseen in-distribution data samples. \\
In the second scenario, we use data from all three numerical simulations, i.e., the TCF at $Re_\tau \approx 1,000$ and the TBL flows at $Re_\tau \approx 390$ and $1,500$, with the same percentage distribution between training, validation, and test samples as for the single configuration ($80$~\%, $10$~\%, $10$~\%). To avoid any bias during training, the same number of samples is taken from each dataset. Thus, $4,164$ samples randomly selected from each dataset are used for training ($12,492$ in total) and $516$ ($1,548$ in total) for validation and testing, respectively.

\section{Results} \label{sec:results} 
In the following, the discussion of the results is split into two main segments. In section~\ref{sec:results1}, we evidence the successful performance of the proposed neural network when trained on a single configuration, i.e., the TCF at $Re_\tau \approx 1,000$. In section~\ref{sec:results2}, we demonstrate that our deep learning architecture accurately predicts wall-shear stress distributions when trained on a comprehensive dataset that unifies flow fields of different configurations, i.e., TCF and TBL flows at three different Reynolds numbers spanning $390 \le Re_\tau \le 1,500$. Moreover, we demonstrate the zero-shot transferability to out-of-distribution experimental data to evidence the strong generalizability and effectiveness in real-world measurements.

\subsection{Wall-shear stress predictions of the turbulent channel flow at $\mathbf{Re_\tau \approx 1,000}$} \label{sec:results1}
This section presents the results of the deep learning framework solely trained on the numerical TCF data at $Re_\tau \approx 1,000$. Section~\ref{subsec:results1_DNS} demonstrates that the network is able to provide accurate in-distribution wall-shear stress predictions based on the numerical test dataset. In section~\ref{subsec:results1_exp}, we show that the network also reliably predicts wall-shear stress distributions from experimental PIV velocity measurements.

\begin{figure}[b!]
\centering
\includegraphics[width=0.65\textwidth]{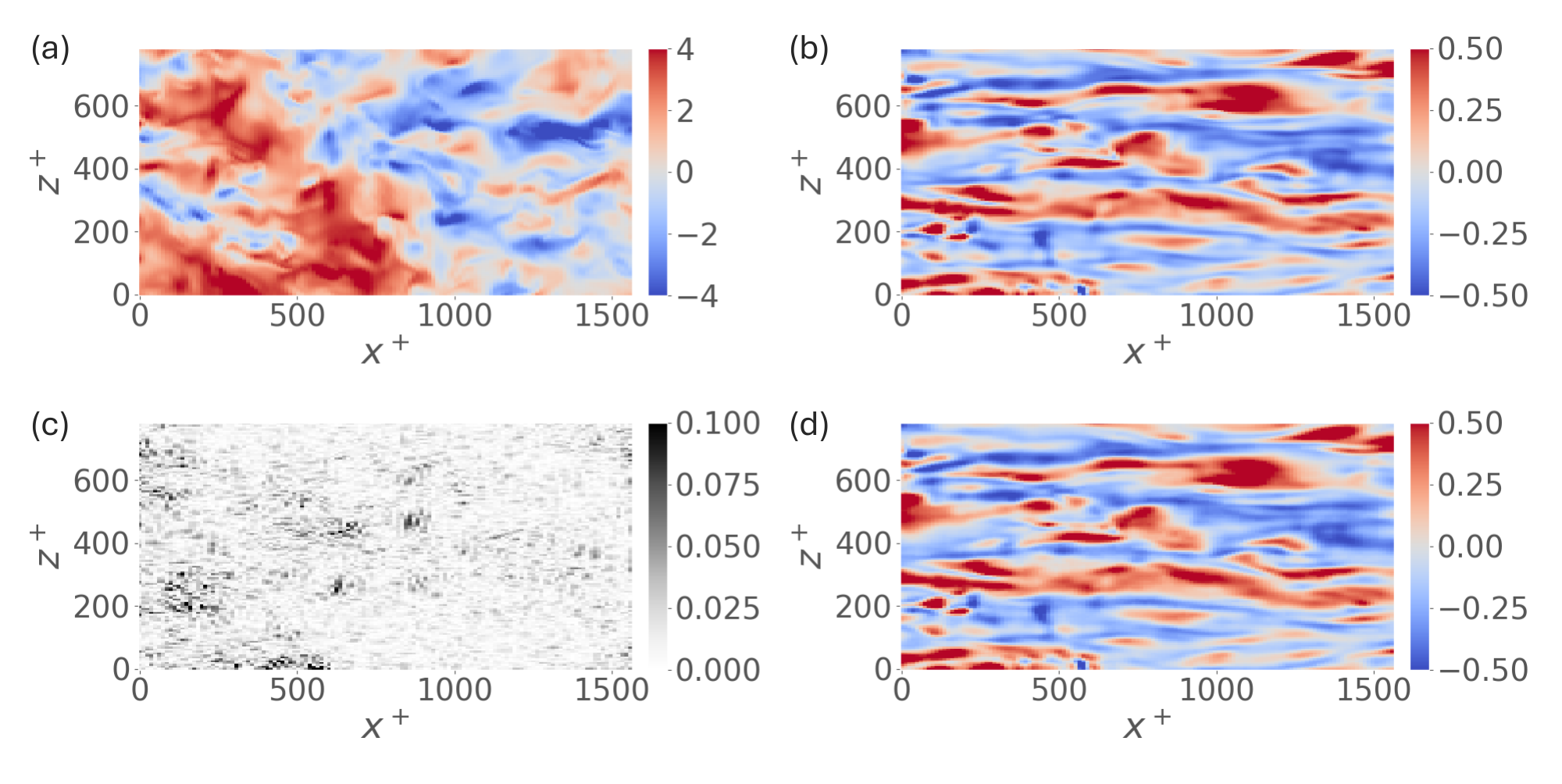}
  \caption{\textbf{Velocity and wall-shear stress fields of an arbitrary sample of the numerical test dataset of the TCF at $\mathbf{Re_\tau \approx 1,000}$.} (a) inner-scaled streamwise velocity fluctuations (network input), (b) true inner-scaled wall-shear stress fluctuations, (c) prediction error, (d) predicted wall-shear stress fluctuations (network output). The comparison between ground truth (b) and predicted (d) wall-shear stress shows that the proposed architecture provides physically correct wall-shear stress fields from unseen velocity fields. This is further evidenced by the low prediction error, which is calculated using the absolute difference between true and predicted wall-shear stress values.}
  \label{fig:instResult_DNS}
\end{figure}

\subsubsection{In-distribution results of the numerical data} \label{subsec:results1_DNS}
Figure~\ref{fig:instResult_DNS} provides an example of an input, i.e., a streamwise velocity fluctuation field, a network output, i.e., the wall-shear stress prediction, the true wall-shear stress distribution, and the error in the network prediction for an arbitrary sample of the test dataset. This example shows how well the network predicts a physically correct wall-shear stress distribution for unseen data (recall that the test dataset was never provided during network training and validation).\\ 

\begin{figure}[b!]
\centering
\includegraphics[width=0.99\textwidth]{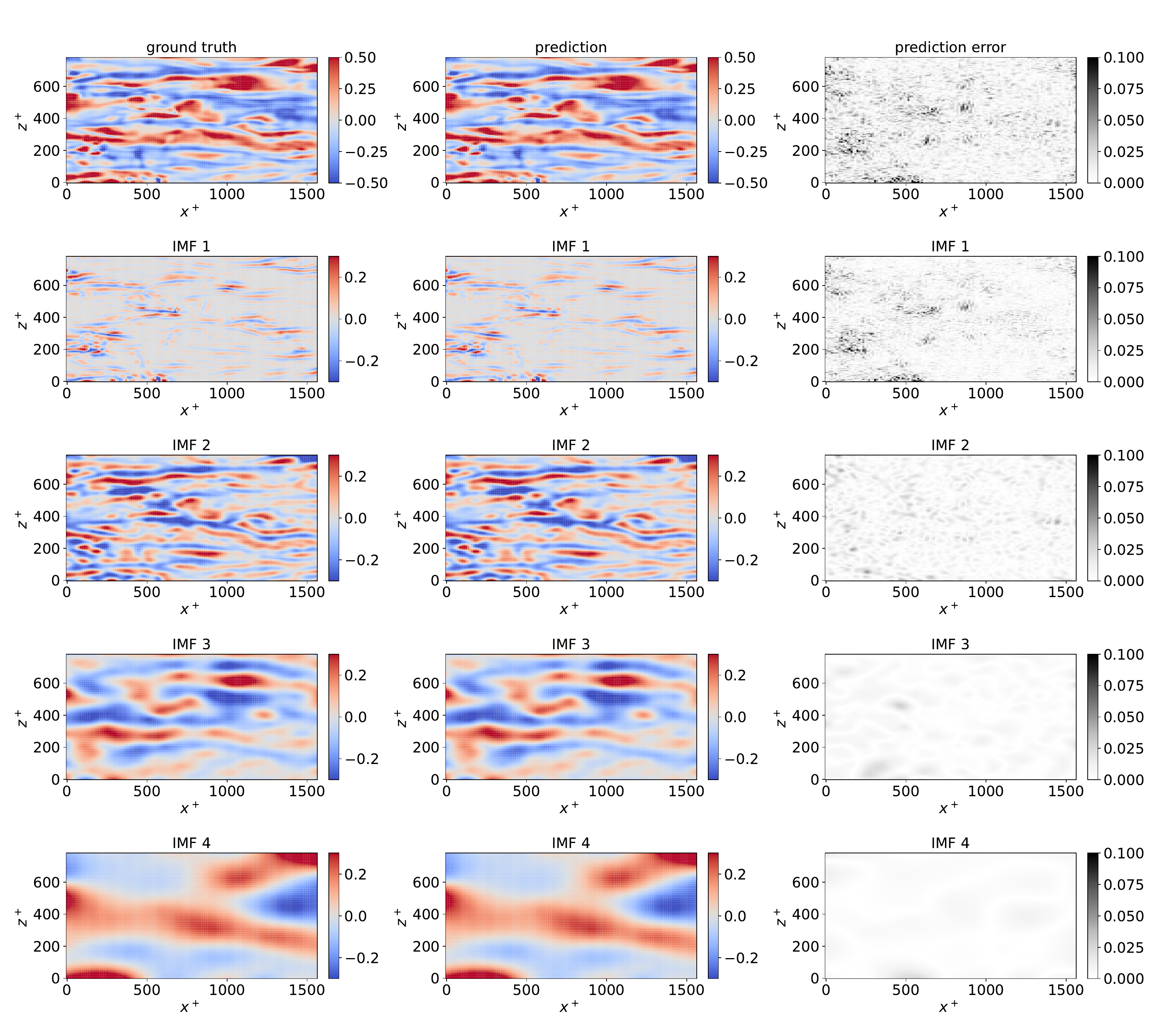}
  \caption{\textbf{Modal representations of the true and the predicted wall-shear stress fields of the test sample presented in figure~\ref{fig:instResult_DNS}.} The left column contains the IMFs of the true wall-shear stress, the center column depicts the IMFs related to the predicted wall-shear stress, and the right column shows the reconstruction error. The first row contains the initial wall-shear stress fields and the corresponding prediction error (prior to applying the 2D NA-MEMD), which are identical to the data in figure~\ref{fig:instResult_DNS}. Although all scales are predicted very accurately by the deep learning framework, the prediction error slightly increases with smaller scale size.}
  \label{fig:instResult_EMD_DNS}
\end{figure}

\begin{figure}[b!]
\centering
\includegraphics[width=0.99\textwidth]{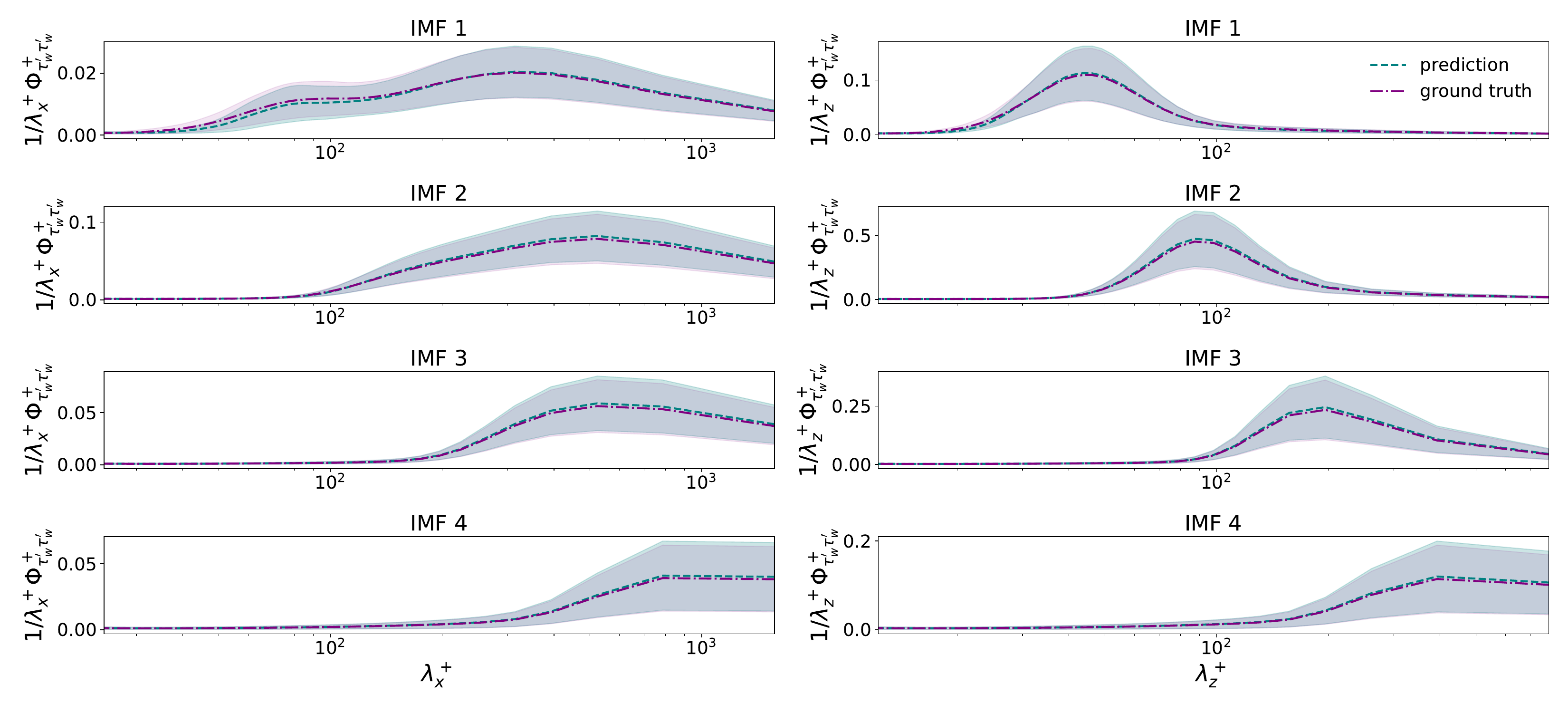}
  \caption{\textbf{Pre-multiplied power spectra of the IMFs of the wall-shear stress predictions and ground truth distributions of the numerical test dataset of the TCF at $\mathbf{Re_\tau \approx 1,000}$.} The left column provides spectra as a function of the streamwise wavelengths $\lambda_x^+$ and the right column as a function of the spanwise wavelengths $\lambda_z^+$. Lines indicate averages over all samples and the shaded regions are the corresponding standard deviations. On a statistical basis, the power spectra evidence how well the network predicts correct wall-shear stress values across all scales. Only a very small deviation can be observed for the smallest scales in IMF1.}
  \label{fig:spectra_EMD_DNS}
\end{figure}

To assess the network's performance in greater detail, we apply the 2D NA-MEMD~\cite{maeteling2022} to the wall-shear stress predictions and ground truth distributions of the test dataset and subsequently perform an FFT on each mode to inspect the power spectra. By simultaneously decomposing the ground truth and the prediction of each sample, we ensure that the resulting IMFs capture the same range of scales. This allows us to assess the prediction capabilities of the network in a scale-based representation. In other words, we can analyze if there is a certain bias in the prediction of specific scales or if a certain range of scales is more difficult to obtain with our deep learning architecture.\\
Figure~\ref{fig:instResult_EMD_DNS} provides an example of the obtained modes of the true and the predicted wall-shear stress distributions of an arbitrary test sample. It is obvious that the largest deviation - which is still small in an overall sense - occurs in the first mode, which comprises the smallest scales. The same tendency can be observed in the inner-scaled pre-multiplied power spectra of the IMFs, which are given in figure~\ref{fig:spectra_EMD_DNS} as a function of the streamwise $\lambda_x^+$ and the spanwise $\lambda_z^+$ wavelengths. The figure depicts the spectra averaged over the entire test dataset and the 
standard deviation across all samples is provided as a shaded area. The distributions prove that the proposed framework provides very accurate wall-shear stress estimates in a statistical sense. Only a small deviation between ground truth and network prediction can be observed in the first mode for small streamwise wavelengths. This behavior is not uncommon for neural networks because they tend to learn large features very well, whereas the small-scale content, which varies most significantly across all samples, is most challenging to be represented reliably. 

In view of the current application, the information about the large-scale wall-shear stress features is already comprised in the velocity field provided as the network input. Extensive studies in the literature~\cite{agostini2014,marusic2007,marusic2017,pathikonda2019,maeteling2022} have shown how the large-scale dynamics of the outer layer are superimposed onto the near-wall flow. Thus, the deep learning framework is essentially a model of the various physical processes involved in this inner-outer interaction. In other words, it can be viewed as a neural operator network, which learns a physically accurate transfer function from the large-scale flow features of the outer-layer velocity field to the wall-shear stress distribution. However, and perhaps surprisingly, figures~\ref{fig:instResult_EMD_DNS},~\ref{fig:spectra_EMD_DNS} show that the proposed architecture is able to learn a physically meaningful representation of the medium- and small-scale content as well (with inaccuracies at the smallest scales, i.e., IMF 1). Most likely, this is rooted in the fact that physical connections between the inner and the outer layer also exist within these scales, but which have been studied less intensely so far because of their less significant impact on the governing near-wall dynamics. However, the smallest scales in turbulent wall-bounded flows have a statistically universal character~\cite{marusic2010_peak}, which means that it is nearly impossible to trace back their origin. Thus, information about the very small-scale behavior of the wall-shear stress cannot be extracted from the outer-layer velocity fields. Therefore, there is a certain inaccuracy in their prediction. However, since these scales have the smallest impact in an energetic sense and are therefore less important for, e.g., drag reduction concepts and life time estimation of structural materials, a minor uncertainty in their correct prediction is acceptable for the majority of applications. 

Thus, in summary, the proposed deep learning architecture performs exceptionally well in predicting wall-shear stress distributions from outer-layer velocity fields for in-distribution numerical data.

\subsubsection{Zero-shot application to experimental data} \label{subsec:results1_exp}
\begin{figure}[t]
\centering
\includegraphics[width=0.99\textwidth]{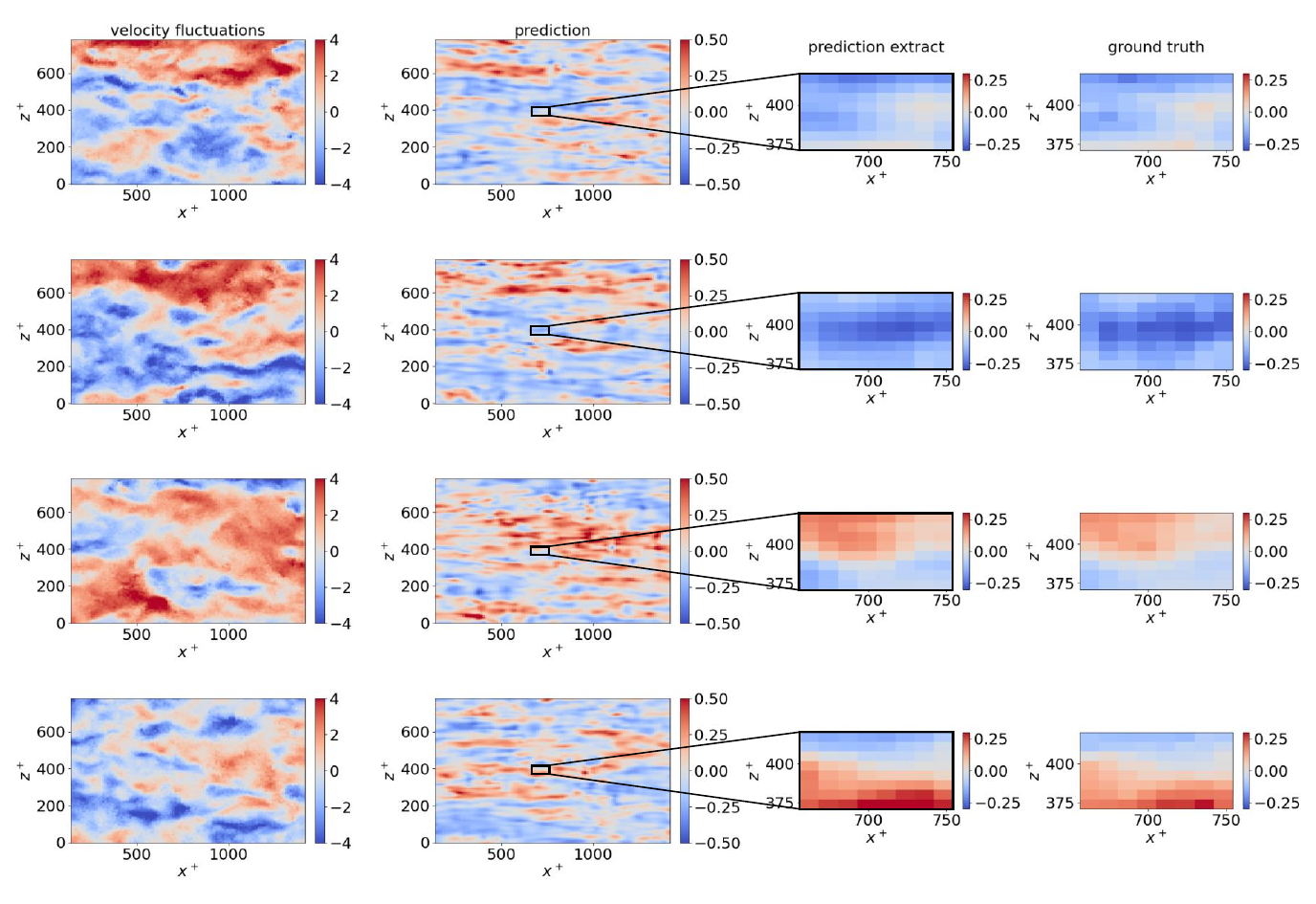}
  \caption{\textbf{Velocity and wall-shear stress fields of arbitrary samples of the experimental dataset of the TCF at $\mathbf{Re_\tau \approx 1,000}$.} The left column shows the inner-scaled streamwise velocity fluctuations obtained from the PIV measurements (network input), the next column depicts the experimental wall-shear stress predictions (network output), and the two right columns present the predicted and MPS$^3$ based wall-shear stress in a spatially limited extract. This comparison shows that the network is able to predict physically meaningful wall-shear stress fields from experimental velocity measurements although it was never trained on the particular characteristics of these measurements.}
  \label{fig:exp_inst}
\end{figure}

The zero-shot applicability of the deep learning framework to experimental data is evidenced with the pre-trained network discussed in section~\ref{subsec:results1_DNS}. 
This neural network is solely trained on DNS data and first exposed to experimental data during testing, i.e., after the training is completed. Figure~\ref{fig:exp_inst} shows the experimental input, i.e., the PIV based velocity fluctuation field, and the predicted wall-shear stress distribution for arbitrary experimental test samples. In addition, the true wall-shear stress distribution measured by the MPS$^3$ is provided next to an enlarged extract of the predicted wall-shear stress field matching the experimentally measured region. The comparison shows a good agreement between prediction and ground truth indicating that the network is able to generalize well to experimental conditions without being explicitly trained on an experimental dataset. Some of the predictions appear smoother compared to the measured wall-shear stress fields due to small-scale measurement uncertainties in the MPS$^3$ data. 

\begin{figure}[t]
\centering
\includegraphics[width=0.99\textwidth]{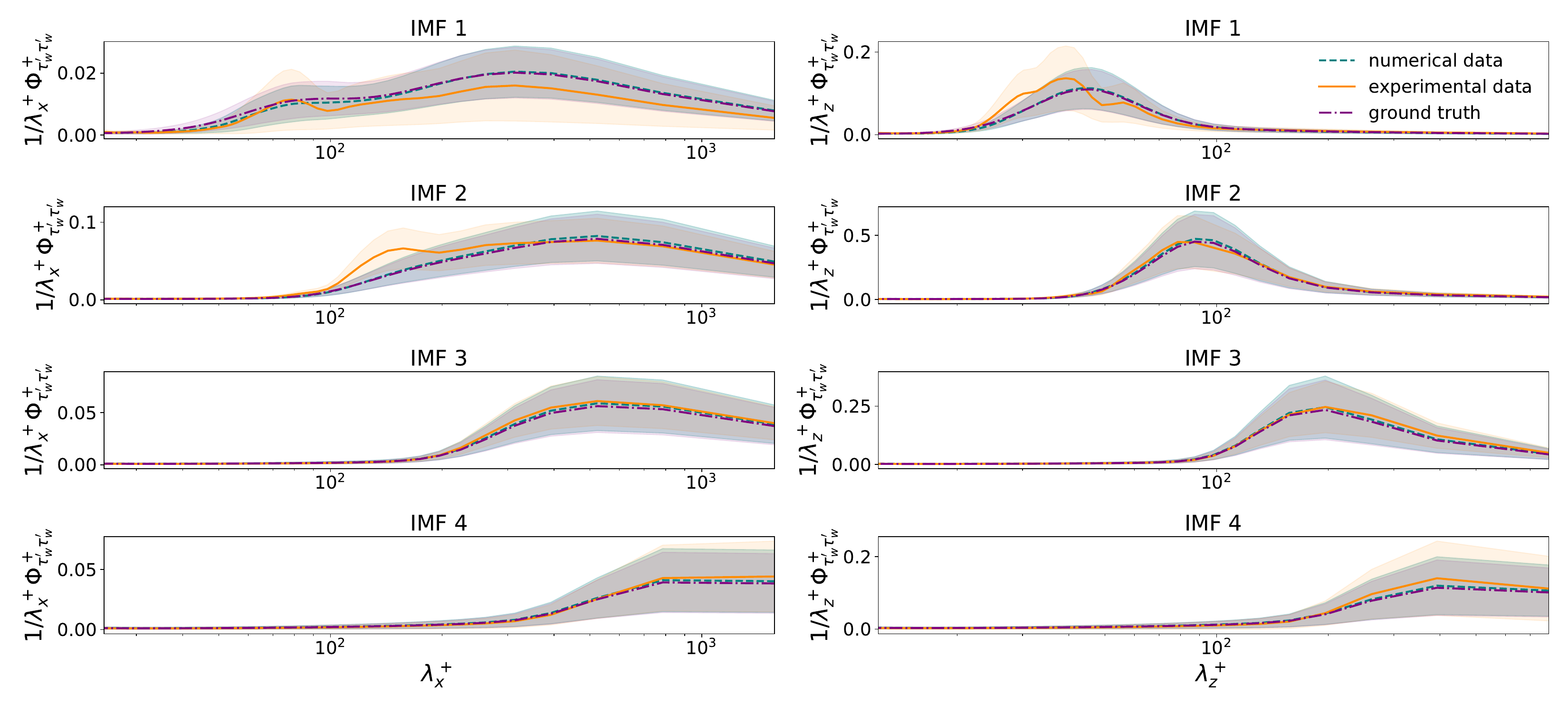}
  \caption{\textbf{Pre-multiplied power spectra of the IMFs of the numerical and experimental wall-shear stress predictions as well as the numerical ground truth distributions of the TCF at $\mathbf{Re_\tau \approx 1,000}$.} The left column provides spectra as a function of the streamwise wavelengths $\lambda_x^+$ and the right column as a function of the spanwise wavelengths $\lambda_z^+$. Lines indicate averages over all samples and the shaded regions are the corresponding standard deviations. Please note that the numerical data were already shown in figure~\ref{fig:spectra_EMD_DNS} and are included for comparison since the statistics of the numerical and the experimental datasets are expected to match. This comparison reveals that the network predictions based on experimental velocity fields are slightly inaccurate with respect to the energetic content of the smallest scales (IMF$1$). This is rooted in the small-scale noise inherent to the experimental velocity measurements (see figure~\ref{fig:OLdistribution}), which is incorrectly processed by the neural network. However, the overall trend of the experimental spectra, especially for medium-size and large-scale flow features, is very similar to the numerical data evidencing a successful performance of the deep learning framework in the noisy flow conditions of the experimental dataset.}
  \label{fig:spectra_EMD_exp}
\end{figure}

To provide further details on the experimental predictions on a statistical basis, especially with respect to medium-size and large-scale flow structures that are not captured in the MPS$^3$ data due to the limited field of view, we apply the 2D NA-MEMD with subsequent FFT to the true numerical, the predicted numerical, and the predicted experimental wall-shear stress fields. Please note that the results related to the numerical data are already given in figure~\ref{fig:spectra_EMD_DNS}. They are included in figure~\ref{fig:spectra_EMD_exp} for comparison since the statistics of the experimental estimates are expected to match the numerical statistics if the deep learning framework performs as intended.
A distinct difference in the power spectra can be observed for the smallest scales, i.e., in IMF1, as well as in the low-wavelength regime of the streamwise spectra of IMF2. The experimental wall-shear stress predictions are not as accurate as the numerical predictions, which results from the small-scale measurement noise inherent to the experimental velocity measurements (see figure~\ref{fig:OLdistribution}). The network is not able to adequately process this information since it was not comprised in the training data distribution, which yields a marginal redistribution of the energetic content. However, we have to acknowledge that these deviations are only within a few percent and the confidence intervals of the numerical and experimental spectra show a substantial overlap. Moreover, the predictions of the medium- and large-scale wavelengths are very accurate. Therefore, we conclude that the proposed architecture is able to estimate physically meaningful wall-shear stress distributions from experimental velocity fields although the training was solely performed with DNS data. It is especially worth mentioning that some of the characteristics of the experimental dataset differ from the numerical dataset as discussed in section~\ref{subsec:Datasets_compare}, with the small-scale measurement noise being the most significant source of variation. Nevertheless, the deep learning framework is able to provide reliable instantaneous (figure~\ref{fig:exp_inst}) and statistical (figure~\ref{fig:spectra_EMD_exp}) information about the experimental wall-shear stress.  

\subsection{Wall-shear stress predictions of the multi-configuration network using turbulent channel and turbulent boundary layer flows at various Reynolds numbers} \label{sec:results2}
In the following, we show that our proposed deep learning architecture predicts reliable wall-shear stress distributions when trained on a comprehensive dataset that unifies flow fields of different configurations. On the one hand, we evidence a successful performance with respect to in-distribution wall-shear stress predictions using the numerical test data. On the other hand, we present a successful zero-shot application to experimental TCFs at several Reynolds numbers. Besides the three Reynolds numbers covered in the training dataset (two of which were related to spatially developing TBL flows and not to fully developed TCF), the network also performs well on an intermediate and an extrapolated Reynolds number.

\begin{figure}[t!]
\centering
\includegraphics[width=0.99\textwidth]{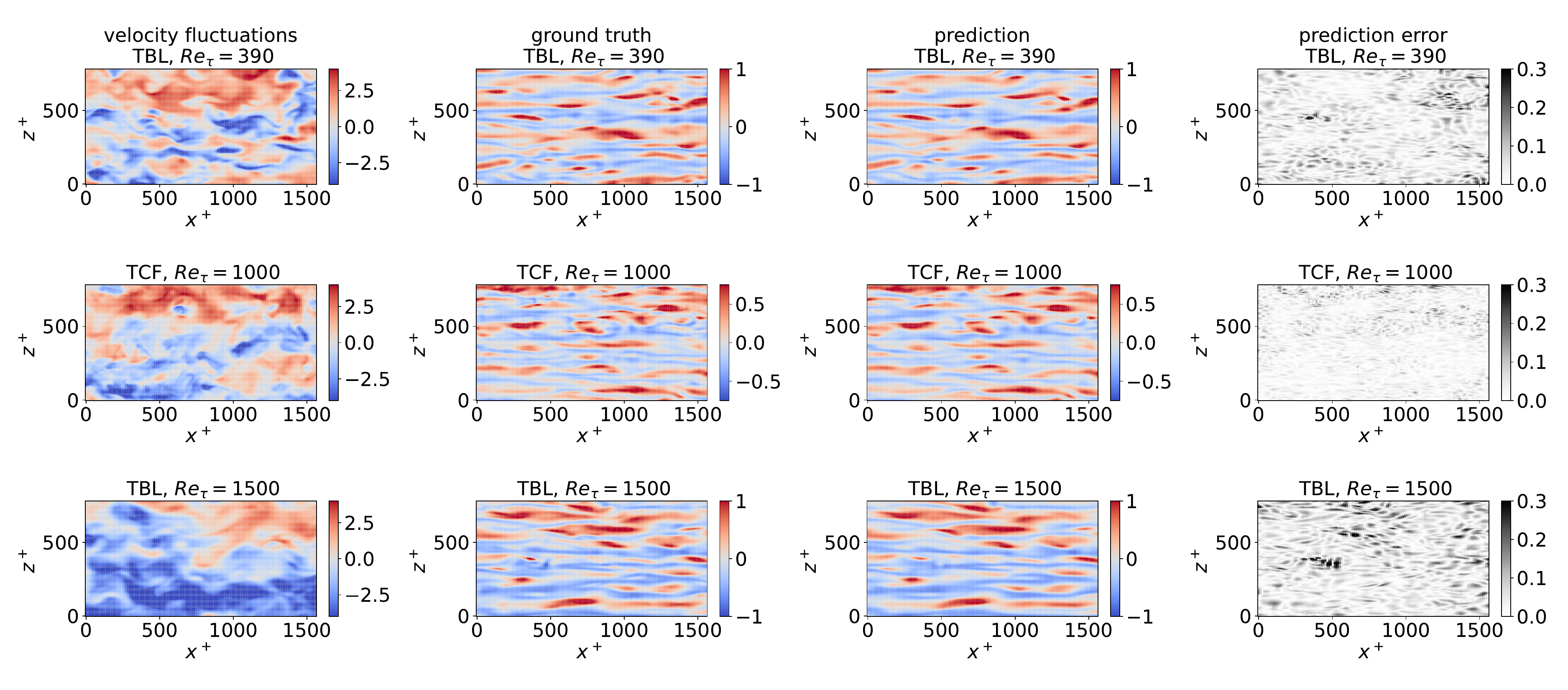}
  \caption{\textbf{Velocity and wall-shear stress fields of arbitrary samples of the numerical test datasets.} From left to right, each row contains the inner-scaled streamwise velocity fluctuations (network input), the true inner-scaled wall-shear stress fluctuations, the predicted wall-shear stress fluctuations (network output), and the prediction error, which is the absolute deviation between ground truth and prediction. From top to bottom, the data correspond to the TBL flow at $Re_\tau \approx 390$, the TCF at $Re_\tau \approx 1,000$, and the TBL flow at $Re_\tau \approx 1,500$. For all configurations, the true and the predicted wall-shear stress distributions match very well and the reconstruction error shows only small-scale irregularities in predicting the true physics (please note the different colorbar limits).}
  \label{fig:instResult_DNS_MCN}
\end{figure}

On an instantaneous level, figure~\ref{fig:instResult_DNS_MCN} shows that the neural network derives accurate wall-shear stress fields for all three configurations. Using arbitrarily selected test samples of the numerical data, each row of figure~\ref{fig:instResult_DNS_MCN} depicts the input velocity field, the true and the predicted wall-shear stress fields, and the reconstruction error for the TBL flow at $Re_\tau \approx 390$ (top), the TCF at $Re_\tau \approx 1,000$ (center), and the TBL flow at $Re_\tau \approx 1,500$ (bottom). The reconstruction errors indicate that the inaccuracy in the prediction is primarily based on small-scale features, which was already observed for the single-configuration setting discussed in section~\ref{sec:results1}. 

\begin{figure}[t!]
\centering
\includegraphics[width=0.99\textwidth]{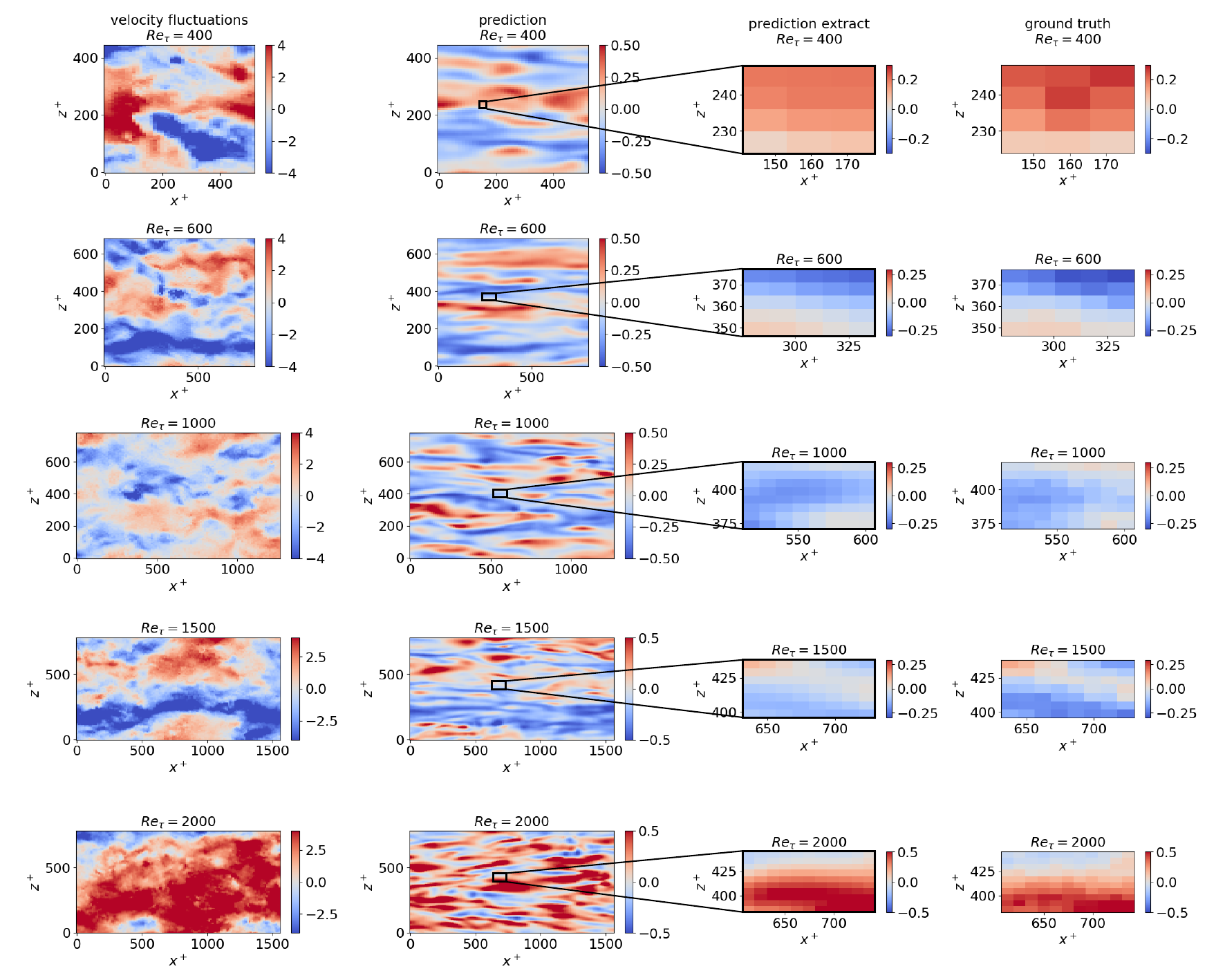}
  \caption{\textbf{Velocity and wall-shear stress fields of arbitrary samples of the experimental TCFs at various Reynolds numbers.} The left column shows the inner-scaled streamwise velocity fluctuations obtained from the PIV measurements (network input), the next column depicts the experimental wall-shear stress predictions (network output), and the two right columns present the predicted and measured wall-shear stress in a spatially limited extract. From top to bottom, the Reynolds number increases from $Re_\tau \approx 400$ to $Re_\tau \approx 2,000$. Even though the training dataset contains only a single numerical configuration of a TCF at $Re_\tau \approx 1,000$, the network also successfully predicts the wall-shear stress fields at the two TBL related Reynolds numbers of $Re_\tau \approx 390$ and $Re_\tau \approx 1,500$. Moreover, the interpolated configuration at $Re_\tau \approx 600$ and the out-of-distribution Reynolds number of $Re_\tau \approx 2,000$ can reliably be covered by the deep learning framework.}
  \label{fig:exp_inst_MCN}
\end{figure}

Figure~\ref{fig:exp_inst_MCN}  presents instantaneous results with respect to the experimental configurations. Each row represents one Reynolds number and displays the PIV based velocity fields, the predicted wall-shear stress fields, and an extract of the former that matches the location and size of the MPS$^3$ based true wall-shear stress distribution. For lower Reynolds numbers, the field of view covered by the PIV measurements is smaller compared to the physical size of the training data. Therefore, the respective velocity and wall-shear stress fields are smaller. Since the deep learning architecture is based on convolutional layers with a small receptive field, a varying input size does not impact the network's performance. \\
The instantaneous results in figure~\ref{fig:exp_inst_MCN} evidence a physically meaningful wall-shear stress prediction across all Reynolds numbers. Although the training data contain solely TBL flows at $Re_\tau \approx 390$ and $Re_\tau \approx 1,500$, the network reliably predicts wall-shear stress fields from TCF related velocity fields at those Reynolds numbers. Moreover, it is able to provide an accurate estimate for a Reynolds number of $Re_\tau \approx 600$, which was not part of the training dataset but within the covered Reynolds number range. Fortunately, we also observe a successful performance at $Re_\tau \approx 2,000$, which constitutes a significant extrapolation from the highest Reynolds number of $Re_\tau \approx 1,500$ included in the training dataset. 

\begin{figure}[t]
\centering
\includegraphics[width=0.99\textwidth]{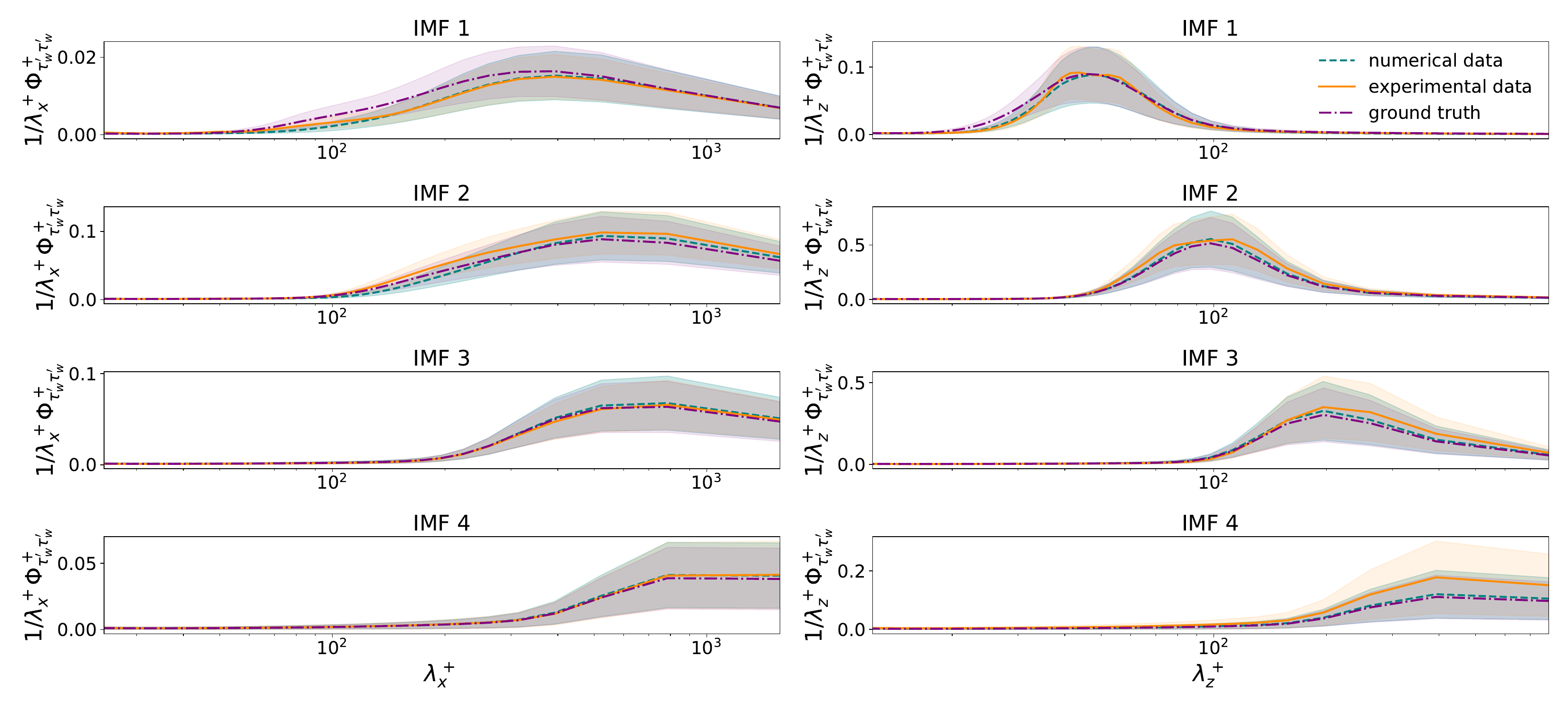}
  \caption{\textbf{Pre-multiplied power spectra of the IMFs of the numerical and experimental wall-shear stress predictions as well as the numerical ground truth distributions at $\mathbf{Re_\tau \approx 390}$.} The left column provides spectra as a function of the streamwise wavelengths $\lambda_x^+$ and the right column as a function of the spanwise wavelengths $\lambda_z^+$. Lines indicate averages over all samples and the shaded regions are the corresponding standard deviations. The predicted wall-shear stress distributions align closely with the ground truth, with the largest discrepancies observed at the smallest scales (IMF1). Generally, the experimental predictions are slightly less accurate than numerical ones, but they still follow the main trends very well. It is important to note that the numerical ground truth and predictions correspond to TBL flow, whereas the experimental data are derived from a TCF, which could result in minor deviations.}
  \label{fig:spectra_EMD_exp_MCN_Re390}
\end{figure}

\begin{figure}[b!]
\centering
\includegraphics[width=0.99\textwidth]{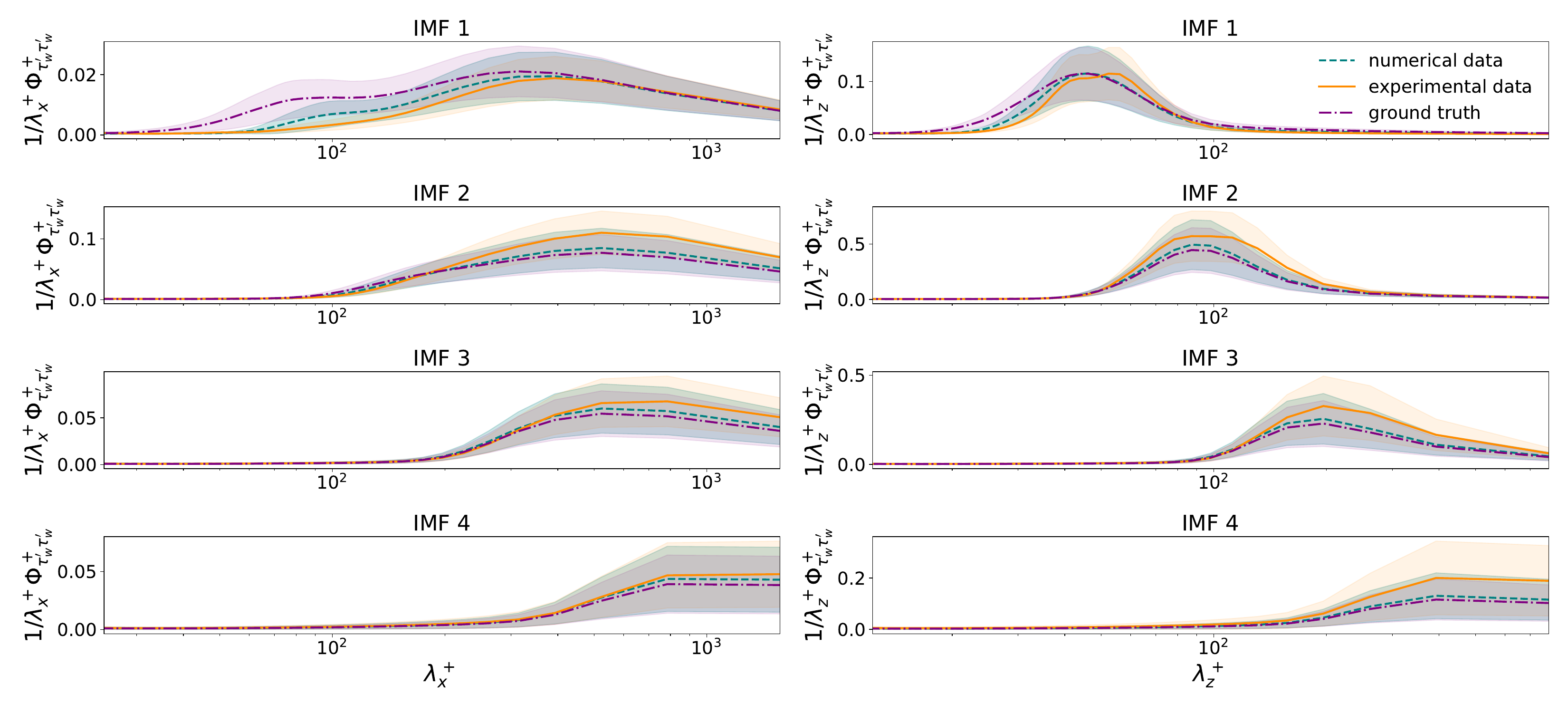}
  \caption{\textbf{Pre-multiplied power spectra of the IMFs of the numerical and experimental wall-shear stress predictions as well as the numerical ground truth distributions at $\mathbf{Re_\tau \approx 1,000}$.} A detailed description is provided in figure~\ref{fig:spectra_EMD_exp_MCN_Re390}.}
  \label{fig:spectra_EMD_exp_MCN_Re1000}
\end{figure}

\begin{figure}[t!]
\centering
\includegraphics[width=0.99\textwidth]{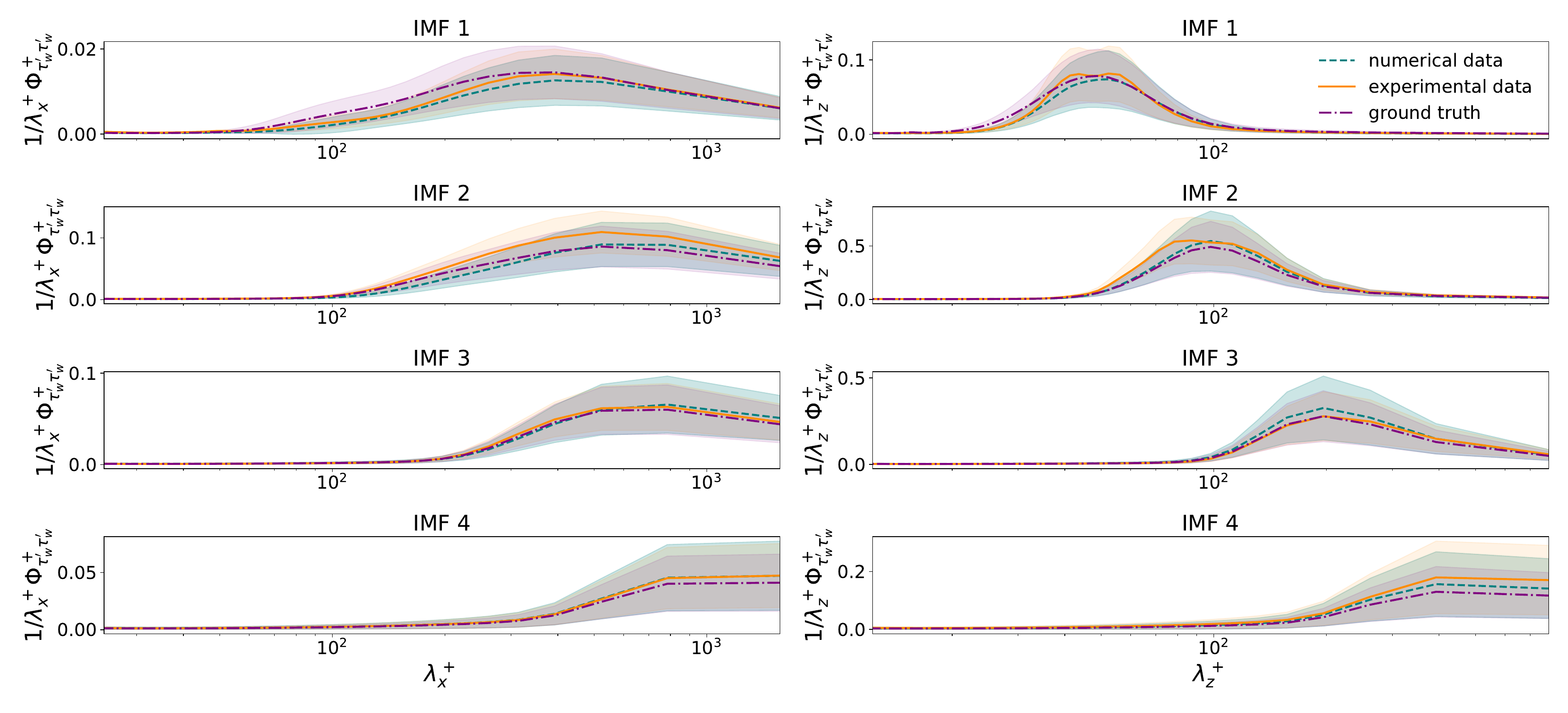}
  \caption{\textbf{Pre-multiplied power spectra of the IMFs of the numerical and experimental wall-shear stress predictions as well as the numerical ground truth distributions at $\mathbf{Re_\tau \approx 1,500}$.} A detailed description is provided in figure~\ref{fig:spectra_EMD_exp_MCN_Re390}.}
  \label{fig:spectra_EMD_exp_MCN_Re1500}
\end{figure}

Next, we investigate the pre-multiplied power spectra obtained by applying an FFT to the 2D NA-MEMD based modal representations of the test data. Figures~\ref{fig:spectra_EMD_exp_MCN_Re390},~\ref{fig:spectra_EMD_exp_MCN_Re1000},~\ref{fig:spectra_EMD_exp_MCN_Re1500} provide the statistics for friction Reynolds numbers of $Re_\tau \approx 390, 1,000$ and $1,500$, respectively. The spectra of the ground truth, the predictions based on the numerical test data, and the predictions based on the experimental data are given as a function of streamwise and spanwise wavelengths. Although the ground truth and the numerical estimates at $Re_\tau \approx 390$ (figure~\ref{fig:spectra_EMD_exp_MCN_Re390}) and $Re_\tau \approx 1500$ (figure~\ref{fig:spectra_EMD_exp_MCN_Re1500}) are based on TBL flows and the experimental data is extracted from TCFs, the statistics are expected to be relatively similar and therefore, presented in one figure. 

Overall, a very good agreement is observed between the ground truth and the numerical predictions for all three Reynolds numbers. Only the smallest scales (IMF1), especially with respect to the streamwise wavelengths, are slightly underestimated. The overall trend of the experimental predictions is very similar to the numerical results but with slightly larger deviations across most modes. Specifically the smallest scales possess a higher deviation from the true distribution, which was equally observed for the single-configuration network discussed in section~\ref{subsec:results1_exp}. By comparing the spectra of the TCF at $Re_\tau \approx 1,000$ predicted with the single-configuration network (figure~\ref{fig:spectra_EMD_exp}) to the multi-configuration network results (figure~\ref{fig:spectra_EMD_exp_MCN_Re1000}), we can see a minor increase of the uncertainty within the prediction when the network is trained on several configurations. This is rooted in the fact that this network has to generalize more broadly across flow conditions, i.e., a range of Reynolds numbers, as well as slightly varying characteristics of the TCF and TBL flows, which comes at the cost of accuracy. Nevertheless, the validation with the measurement data on an instantaneous level (figure~\ref{fig:exp_inst_MCN}) as well as the power spectra still show a reliable prediction and generalization ability that allows a trustworthy application to experimental data covering a broad range of Reynolds numbers.

\begin{figure}[b!]
\centering
\includegraphics[width=0.99\textwidth]{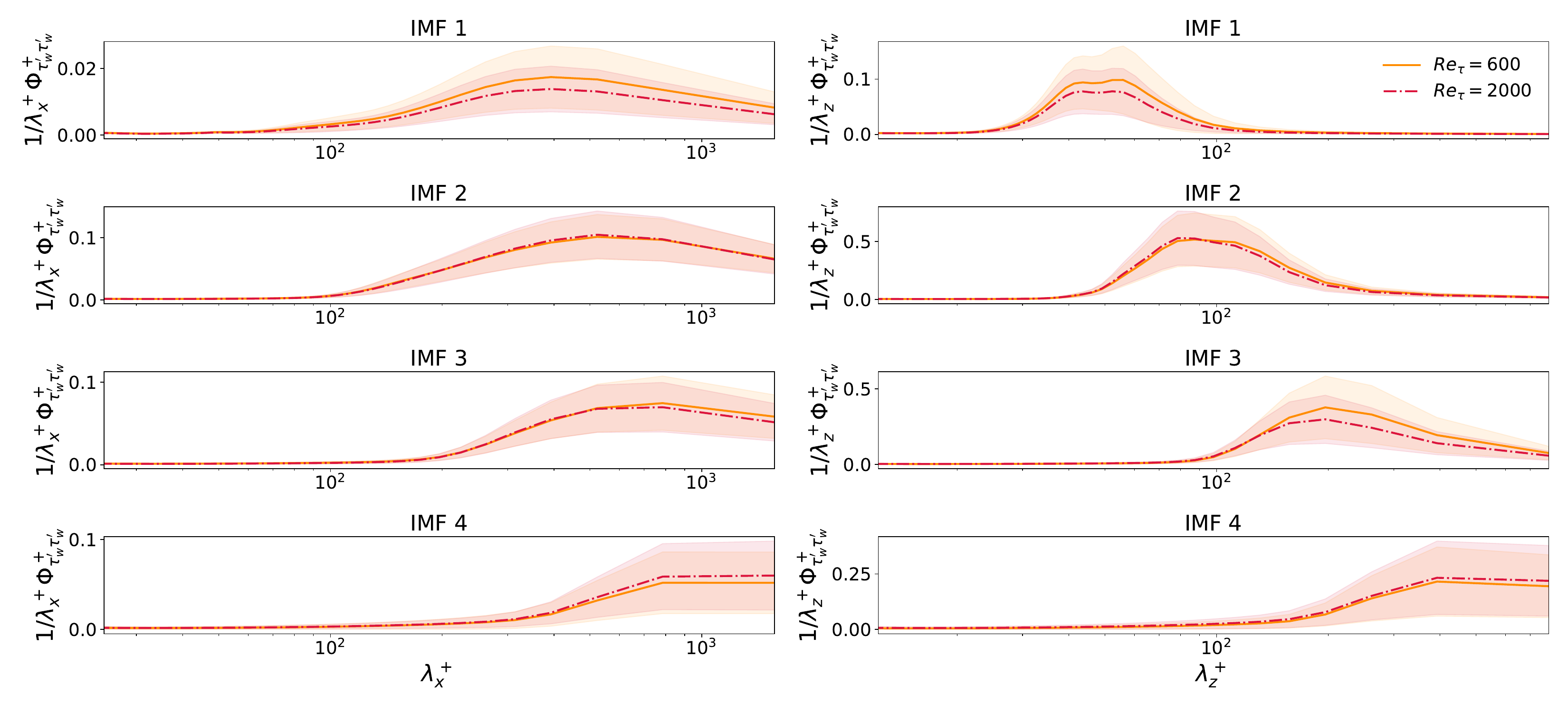}
  \caption{\textbf{Pre-multiplied power spectra of the IMFs of the experimental wall-shear stress predictions at $\mathbf{Re_\tau \approx 600}$ and $\mathbf{Re_\tau \approx 2,000}$.} The left column provides spectra as a function of the streamwise wavelengths $\lambda_x^+$ and the right column as a function of the spanwise wavelengths $\lambda_z^+$. Lines indicate averages over all samples and the shaded regions are the corresponding standard deviations. Although no ground truth information is available for comparison, the overall trends of these spectra are physically meaningful and similar to the other Reynolds numbers shown in figures~\ref{fig:spectra_EMD_exp_MCN_Re390},~\ref{fig:spectra_EMD_exp_MCN_Re1000},~\ref{fig:spectra_EMD_exp_MCN_Re1500}. Indeed, for these Reynolds numbers and the depicted range of wavelengths, the inner-scaled pre-multiplied power spectra are expected to collapse~\cite{samie2018fully}.}
  \label{fig:spectra_EMD_exp_MCN_600_2000}
\end{figure}

Lastly, we analyze the statistics at Reynolds numbers which were not covered by the training data. Precisely, the power spectra of the TCF at $Re_\tau \approx 600$ (within the covered range) and at $Re_\tau \approx 2,000$ (extrapolation) are presented in figure~\ref{fig:spectra_EMD_exp_MCN_600_2000}. Their distributions are very similar to the other Reynolds numbers shown in figures~\ref{fig:spectra_EMD_exp_MCN_Re390},~\ref{fig:spectra_EMD_exp_MCN_Re1000},~\ref{fig:spectra_EMD_exp_MCN_Re1500} with the largest deviation occurring in IMF1. Although no ground truth information is available for a thorough comparison, the power spectra of all considered Reynolds numbers are expected to be very similar in the investigated range of wavelengths~\cite{samie2018fully}. Therefore, and in view of the good agreement with the instantaneous wall-shear stress measurements, we conclude that the network is able to predict reliable wall-shear stress information for these Reynolds numbers. This ability is a very useful characteristic for prospective applications since the neural architecture does not need to be trained on the exact flow conditions targeted in the experimental measurements. By covering a sufficiently large range of Reynolds numbers and flow problems with enough intermediate stages, our study indicates that interpolation and (reasonable) extrapolation to additional flow configurations is possible. Thus, the proposed neural architecture provides a valuable surrogate model for temporally and spatially resolved wall-shear stress dynamics captured within a large spatial domain to advance future experimental applications.  

\section{Conclusion} \label{sec:discussion}
Acquiring experimental wall-shear stress information is of tremendous value for academic and applied research in a variety of disciplines ranging from human medicine to civil aviation. However, experimental measurements of temporally and spatially resolved wall-shear stress dynamics covering a large spatial domain are extraordinarily challenging - often even impossible. In the present work, we provide a deep learning based surrogate model which derives instantaneous wall-shear stress distributions from wall-parallel velocity fields located within the logarithmic region. In many settings, velocity measurements in this region are easily conducted using planar Particle-Image Velocimetry (PIV). Purely trained on Direct Numerical Simulation (DNS) data, our proposed neural architecture outputs reliable wall-shear stress fields from PIV-based velocity fields. We verify the physical significance of these neural predictions with wall-shear stress measurements, which have been synchronized to the PIV measurements, using the Micro-Pillar Shear-Stress Sensor (MPS$^3$). 

Our study is based on fully developed turbulent channel flows (TCFs) at friction Reynolds numbers in the range of $400 \le Re_\tau \le 2,000$ as well as spatially developing turbulent boundary layer (TBL) flows at $Re_\tau \approx 390$ and $Re_\tau \approx 1,500$. The velocity fields are extracted in a wall-parallel plane located at $y^+ \approx 3.9 \sqrt{Re_\tau}$ because the large-scale flow features are most energetic in this region. Thus, they have the most significant impact on the wall-shear stress dynamics due to the various physical mechanisms involved in the inner-outer interaction inherent to turbulent wall-bounded flows. Using a supervised training approach based on DNS data, the proposed deep learning architecture learns a neural representation of these interaction phenomena such that it is able to extract physically correct wall-shear stress distributions from the provided velocity fields. We demonstrate the successful performance of this framework when solely trained on a single flow problem, i.e., the TCF at $Re_\tau \approx 1,000$, as well as a multi-configuration setting in which TCF and TBL flows at several Reynolds numbers are combined. Furthermore, our deep learning framework demonstrates zero-shot applicability to experimental measurement data of TCFs. Specifically, it provides physically accurate wall-shear stress information from PIV-based velocity fields at various Reynolds numbers without being exposed to the particular characteristics of the measurement data during training. Notably, the framework has proven its predictive capabilities even for Reynolds numbers beyond the scope of the training dataset, including one intermediate and one extrapolated flow condition. This behavior underscores the robustness of our deep learning architecture and its ability to provide accurate wall-shear stress estimates without the necessity of training on the exact flow conditions, which represents a substantial leap towards achieving true generalizability in fluid dynamics modeling. \\ 
The results of this study bear great potential for experimental and numerical fluid dynamics. Being able to use the proposed framework as a surrogate model to obtain experimental wall-shear stress dynamics can advance research in various domains. For example, it allows to experimentally study the efficiency and the involved physical mechanisms of friction drag reduction techniques, which are mainly investigated by numerical methods so far. In this context, it might also be used in experimental feedback control applications where only access to velocity data is available. On the other hand, our framework can advance the experimental in-vitro investigation of cardiovascular diseases, in which the temporal dynamics as well as the spatial distribution of the wall-shear stress play a central role in disease development and progression. With respect to numerical simulations, our workflow could support the development of enhanced or novel wall models for wall-modeled large-eddy simulations since it constitutes a purely data-driven representation of the inner-outer interactions, which could generalize more broadly than state-of-the-art analytical models when trained on a variety of different flow conditions.\\
To enable these prospective advancements, future studies have to focus on the generalization capabilities in a broader setting. So far, we considered two flow problems at various Reynolds numbers, but the physical mechanisms of the inner-outer interactions within TCFs and TBL flows are quite similar. Thus, an extension to a broader range of flow problems with a higher variability in the respective flow conditions is required. Moreover, it is certainly of great interest to understand how the neural network transfers outer-layer velocity to wall-shear stress information. Therefore, the application of symbolic regression~\cite{brunton2016,udrescu2020,udrescu2020b} to the latent space could provide an analytical expression of the inherent mapping function, which can subsequently be analyzed in a physical context and used for model-based predictive tasks.

\textbf{Acknowledgments:}\\
This research was funded by the German Research Foundation within the Walter Benjamin fellowships MA~10764/1-1 (EL) and LA~5508/1-1 (CL). EL, CL, and SLB acknowledge support from the National Science Foundation AI Institute in Dynamic Systems (grant number 2112085) and the Boeing Company. Furthermore, the authors gratefully acknowledge the Gauss Centre for Supercomputing e.V. for funding this project by providing computing time on the GCS Supercomputers. Finally, we like to thank Wolfgang Schröder and the Institute of Aerodynamics of RWTH Aachen University for supporting the experimental measurement campaign.

\bibliographystyle{plain}

\newpage
\appendix
\counterwithin{figure}{section}

\section{Wall-shear stress predictions with residual and Unet-based architectures} \label{sec:appendix}
In this section, we present results of the single-configuration setup, i.e., using only the TCF at $Re_\tau \approx 1,000$, for two alternative architectures. These are based on the famous residual neural network (ResNet)~\cite{he2016} and the Unet~\cite{ronneberger2015}. After introducing the architectures in section~\ref{appd:architecture}, in-distribution results with respect to the numerical test data are presented in section~\ref{appd:numerical}. The performance on experimental measurement data is discussed in section~\ref{appd:experimental}.

\subsection{Architectural details} \label{appd:architecture}
\begin{figure}[h]
\centering
\includegraphics[width=0.9\textwidth]{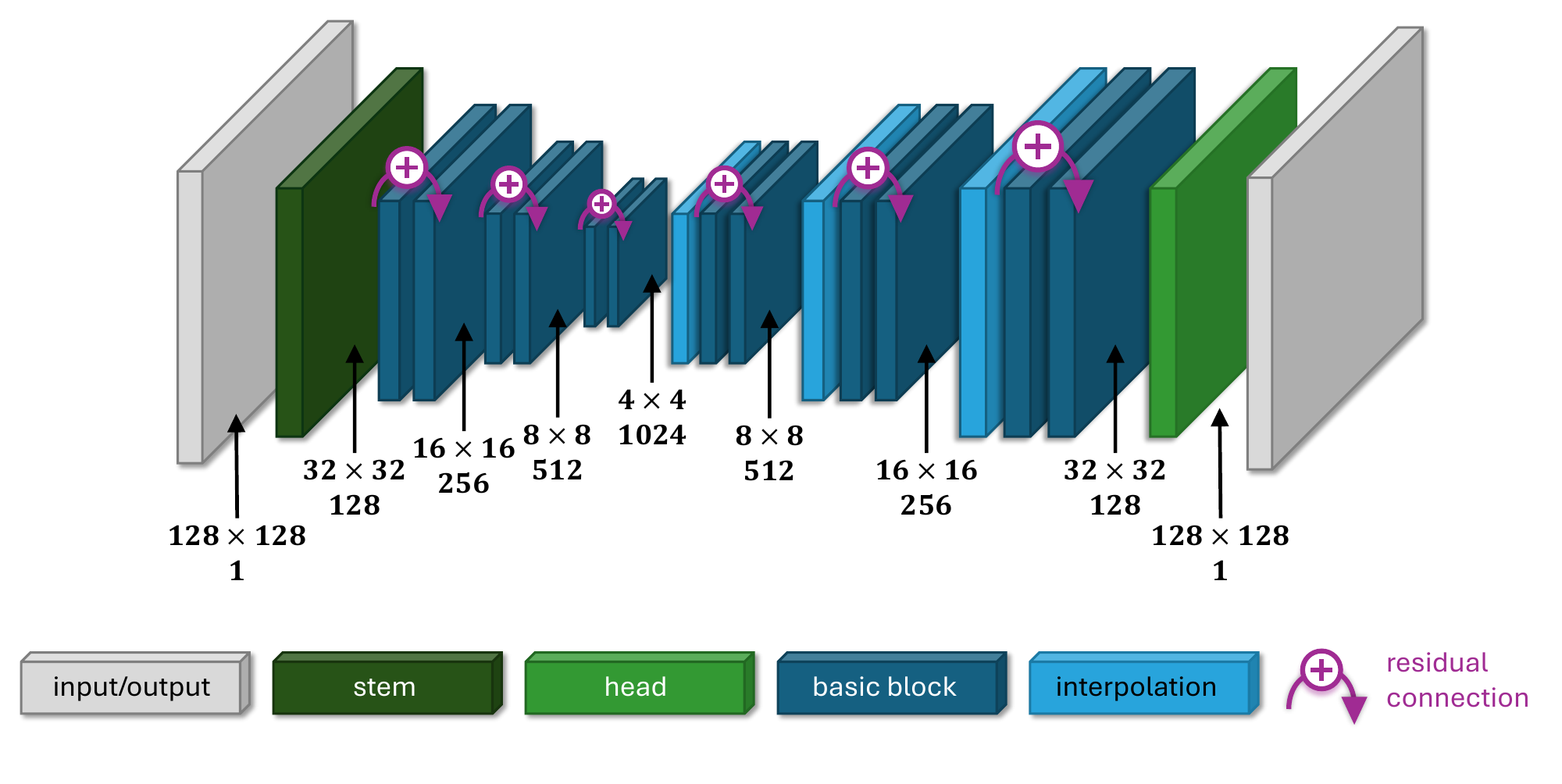}
  \caption{\textbf{Sketch of the ResNet architecture.} The network composition is identical to the original setup given in figure~\ref{fig:architecture} except for residual connections across the basic modules. Through these skip connections, the input to the respective module is added to its output. If a module comprises more than two basic blocks - recall that the number of blocks per module is a hyperparameter -, residual connections are applied between sets of two consecutive blocks. Further details on the architecture and the abbreviations are given in figure~\ref{fig:architecture}.}
  \label{fig:architecture_RES}
\end{figure}

\begin{figure}[h]
\centering
\includegraphics[width=0.9\textwidth]{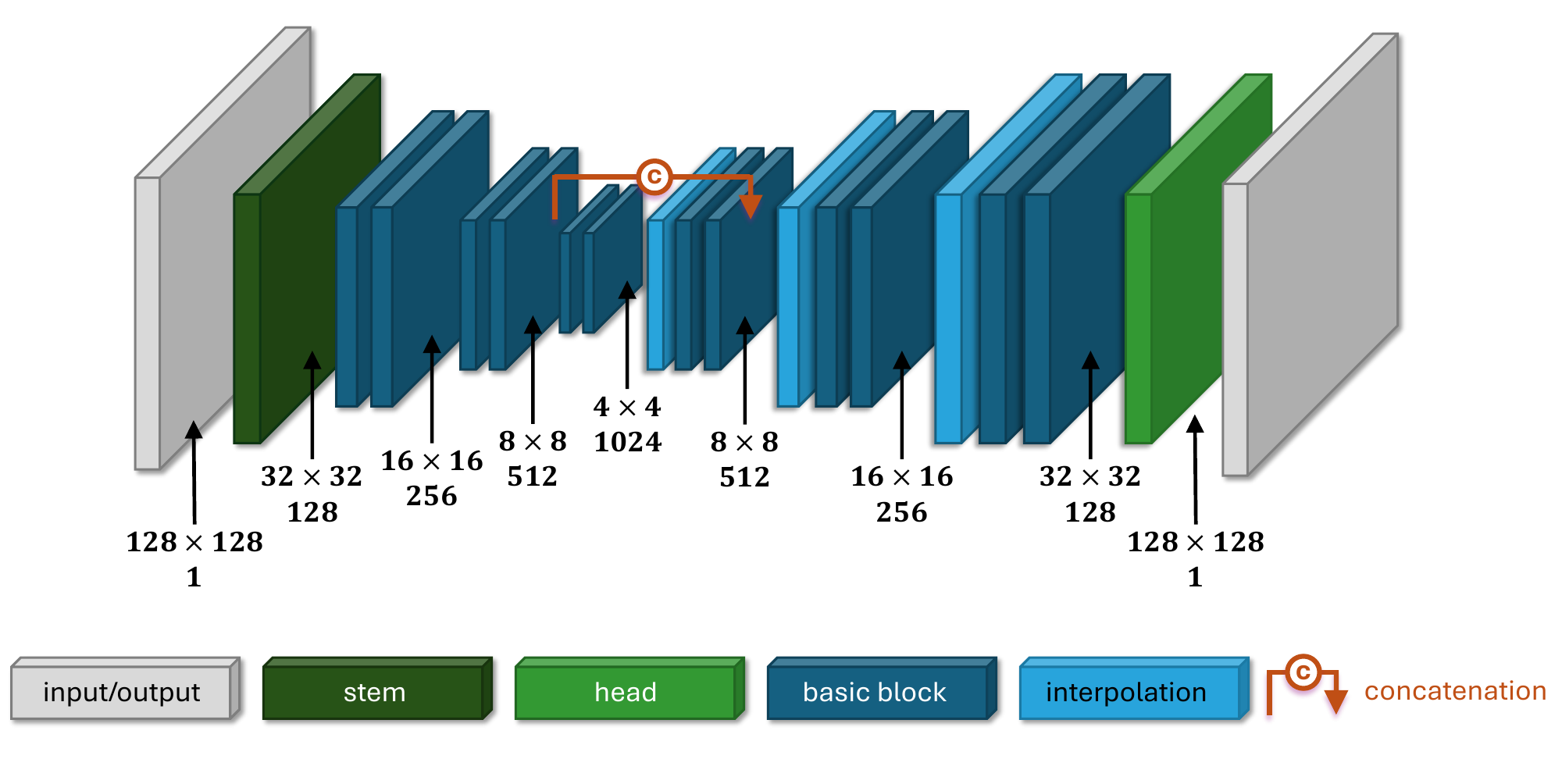}
  \caption{\textbf{Sketch of the Unet architecture.} The network composition is identical to the original setup given in figure~\ref{fig:architecture} except for an additional connection between the last encoding and the first decoding module ("concatenation"). Precisely, the input to the last encoding module is concatenated to the output of the first decoding module. Further details on the architecture and the abbreviations are given in figure~\ref{fig:architecture}.}
  \label{fig:architecture_Unet}
\end{figure}

The fundamental architectures of the residual network (ResNet) and the Unet are very similar to the architecture introduced in section~\ref{sec:architecture} and sketched in figures~\ref{fig:architecture_RES},~\ref{fig:architecture_Unet}, respectively. For the ResNet, a residual connection (or skip connection) is applied to each basic module meaning that the input of this module is added to the output. Architectures based on these residual blocks can overcome the vanishing gradient problem in deep neural networks and thus, potentially result in a better performance~\cite{he2016}. Using hyperparameter studies, we optimized the number of basic blocks in each module as well as the training parameters. The best performance is achieved with an encoder consisting of four basic blocks in the first two basic modules and two basic blocks in the third module and a decoder with two basic blocks per module. We used a batch size of $7$ and a dropout ratio of $0.3$ to prevent overfitting. An Adam optimizer with an initial learning rate of $9.7 \cdot 10^{-5}$ and weight decay of $1.7 \cdot 10^{-4}$ was combined with a scheduler that reduces the learning rate by a factor of $0.2$ when the validation loss is not improving after $10$ epochs. In total, training was performed for $500$ epochs with a minimum learning rate of $10^{-10}$. 

The proposed Unet architecture has a single cross-connection between the last encoding module, i.e., the stage prior to the most compressed layer, and the output of the first decoding module (see figure~\ref{fig:architecture_Unet}). By concatenating the encoded information to the decoding path, a better network performance was observed in other studies~\cite{ronneberger2015}. However, in contrast to classical autoencoder applications, we are not trying to replicate the input data. Thus, concatenating high-resolution information of the encoding path to the decoder at less compressed stages has actually an adverse effect on the network's performance, which was verified in our preliminary studies (not shown for brevity). The present setup requires less encoding and decoding layers compared to the other architectures, i.e., only two basic blocks in each module. Therefore, the total number of trainable parameters is reduced. For training, the same training parameters, which have been used for the ResNet, are applied.

\subsection{In-distribution results of the numerical data} \label{appd:numerical}
\begin{figure}[h]
\centering
\includegraphics[width=0.99\textwidth]{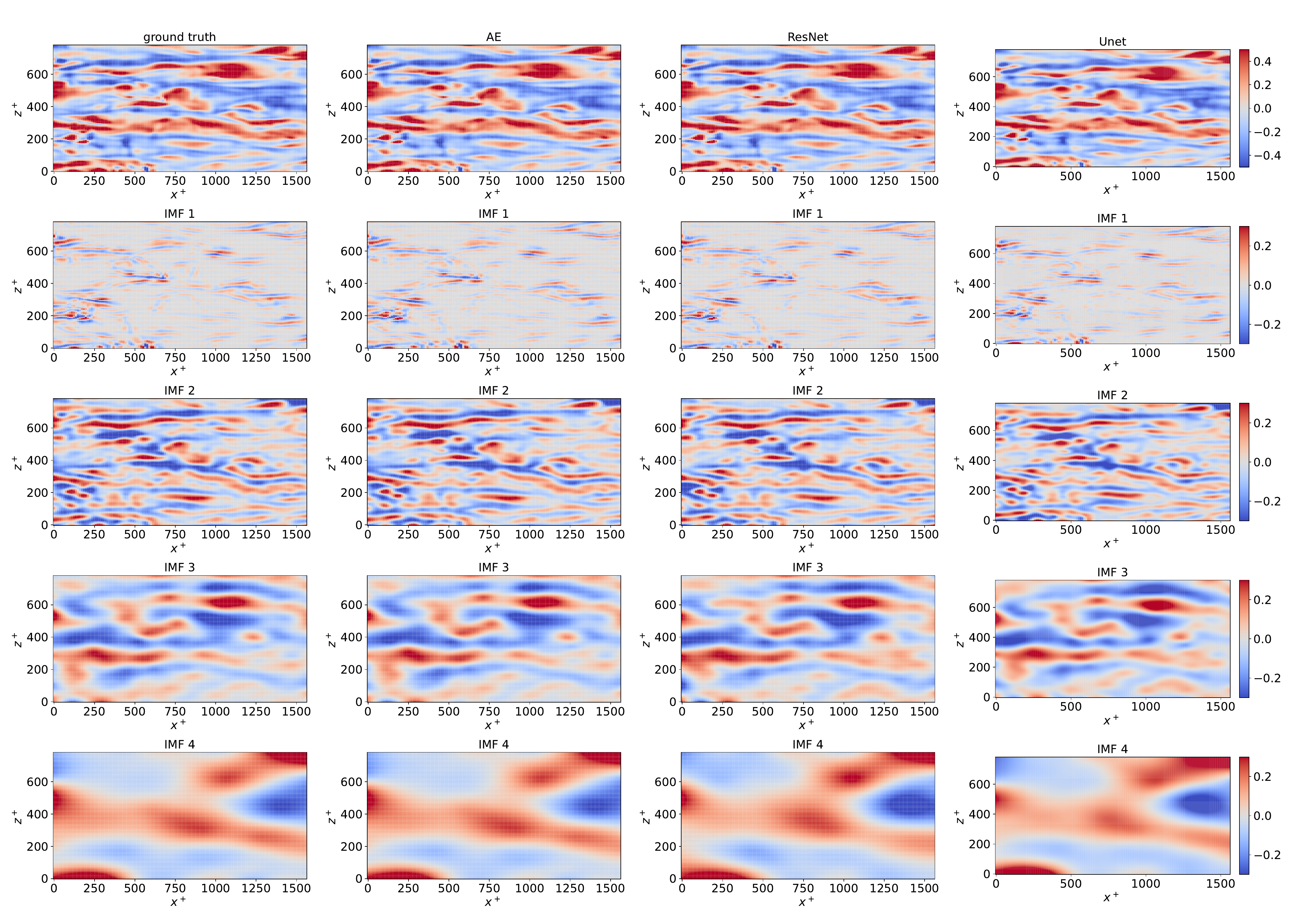}
  \caption{\textbf{Modal representations of the true and the predicted wall-shear stress fields of an arbitrary numerical test sample of the TCF at $\mathbf{Re_\tau \approx 1,000}$ for all three architectures.} The left column contains the IMFs of the true wall-shear stress and the other columns depict the IMFs related to the predicted wall-shear stress (from left to right: basic original autoencoder (AE), ResNet, Unet). The first row contains the initial wall-shear stress fields prior to applying the 2D NA-MEMD, while the remaining rows (from top to bottom) depict modes of increasing scale size. Overall, all three architectures perform similarly well and no distinct difference can be observed with respect to the ground truth fields.}
  \label{fig:imfs_all}
\end{figure}

In figure~\ref{fig:imfs_all}, instantaneous results of an arbitrary sample of the numerical test dataset are shown for all three architectures. In addition to the predicted wall-shear stress fields and the ground truth, the four modal representations are provided. That is, each column represents one configuration (from left to right: ground truth, original architecture (AE), ResNet, Unet), the upper row depicts the total wall-shear stress fields, and the other rows are IMFs (increasing mode number from top to bottom). Overall, the predictions as well as the individual scale-based modal representations are very similar and match the ground truth nearly perfectly. Thus, on an instantaneous level, no significant difference can be observed between the three architectures. 

\begin{figure}[h]
\centering
\includegraphics[width=0.99\textwidth]{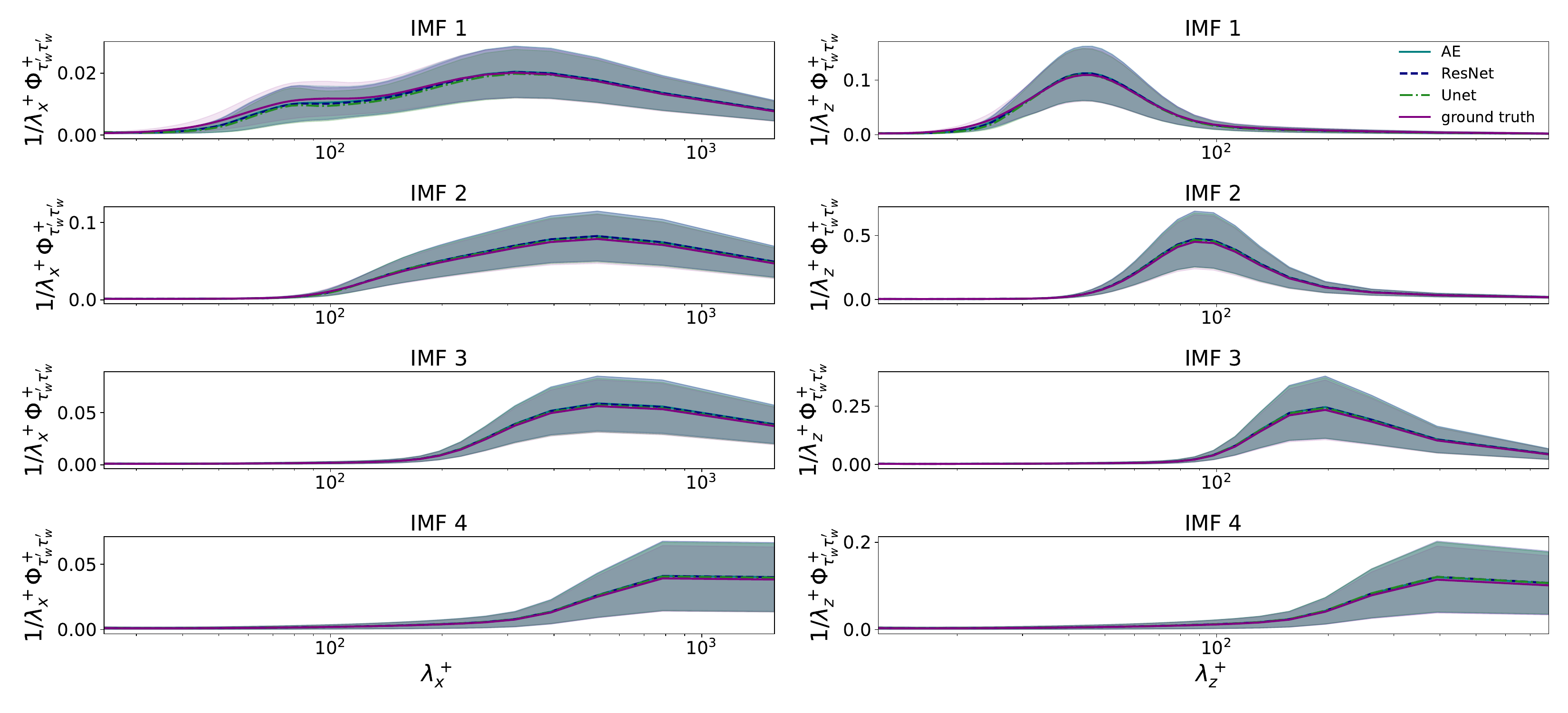}
  \caption{\textbf{Pre-multiplied power spectra of the IMFs of the numerical wall-shear stress predictions as well as the numerical ground truth distributions of the TCF at $\mathbf{Re_\tau \approx 1,000}$.} The left column provides spectra as a function of the streamwise wavelengths $\lambda_x^+$ and the right column as a function of the spanwise wavelengths $\lambda_z^+$. Lines indicate averages over all samples and the shaded regions are the corresponding standard deviations. No major differences are observed for the different architectures and all spectra follow the ground truth very well.}
  \label{fig:spectra_EMD_DNS_all}
\end{figure}

To inspect potential differences in a statistical sense, figure~\ref{fig:spectra_EMD_DNS_all} provides IMF based inner-scaled pre-multiplied power spectra for all three architectures and the ground truth. The left column depicts spectra as a function of the streamwise wavelengths and the right column as a function of the spanwise wavelengths. Except for very small deviations at the smallest scales represented by IMF1, no difference can be observed between the investigated architectures. Therefore, we conclude that for in-distribution numerical data, all networks perform similarly well in reconstructing the wall-shear stress fields.

\subsection{Zero-shot application to experimental data} \label{appd:experimental}
\begin{figure}[t]
\centering
\includegraphics[width=0.99\textwidth]{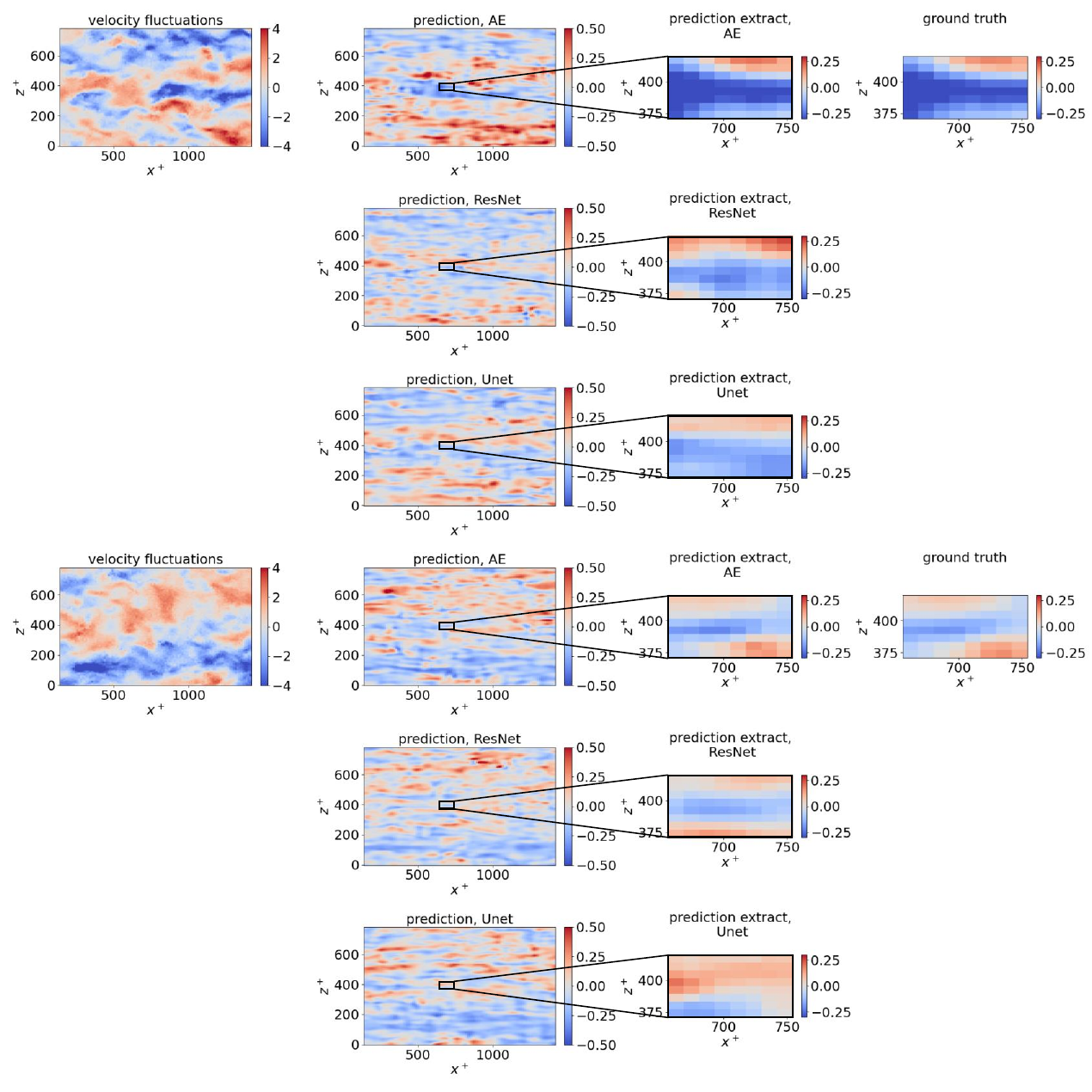}
  \caption{\textbf{Velocity and wall-shear stress fields of arbitrary samples of the experimental dataset of the TCF at $\mathbf{Re_\tau \approx 1,000}$.} The left column shows the inner-scaled streamwise velocity fluctuations obtained from the PIV measurements (network input), the next column depicts the experimental wall-shear stress predictions (network output), and the two right columns present the predicted and MPS$^3$ based wall-shear stress in a spatially limited extract. The original architecture (AE) performs best in an instantaneous fashion as the predictions are closest to the ground truth distributions. Nevertheless, the predictions of the governing dynamics are similar for all three networks.}
  \label{fig:exp_inst_all}
\end{figure}

To inspect the different networks' ability to handle experimental data, the PIV-based velocity measurements of the TCF at $Re_\tau \approx 1,000$ are processed by all three architectures. Examples of these input fields as well as of the instantaneous wall-shear stress predictions are given in figure~\ref{fig:exp_inst_all}. Moreover, the MPS$^3$ based true wall-shear stress distributions are provided next to equally positioned extracts of the wall-shear stress predictions. \\
The comparison to the true wall-shear stress distributions reveals that the original architecture (AE) most reliably predicts accurate wall-shear stress information. Especially with respect to the upper sample, the AE estimates flow features of higher magnitude and thus, also predicts stronger gradients. In contrast, the ResNet and the Unet based architectures predict a smoother distribution with less intense fluctuations. For the provided samples, the ResNet's performance is slightly better than the Unet's ability to predict accurate wall-shear stress fields. However, the governing (large-scale) flow features are similarly well predicted with all three architectures. 

\begin{figure}[t]
\centering
\includegraphics[width=0.99\textwidth]{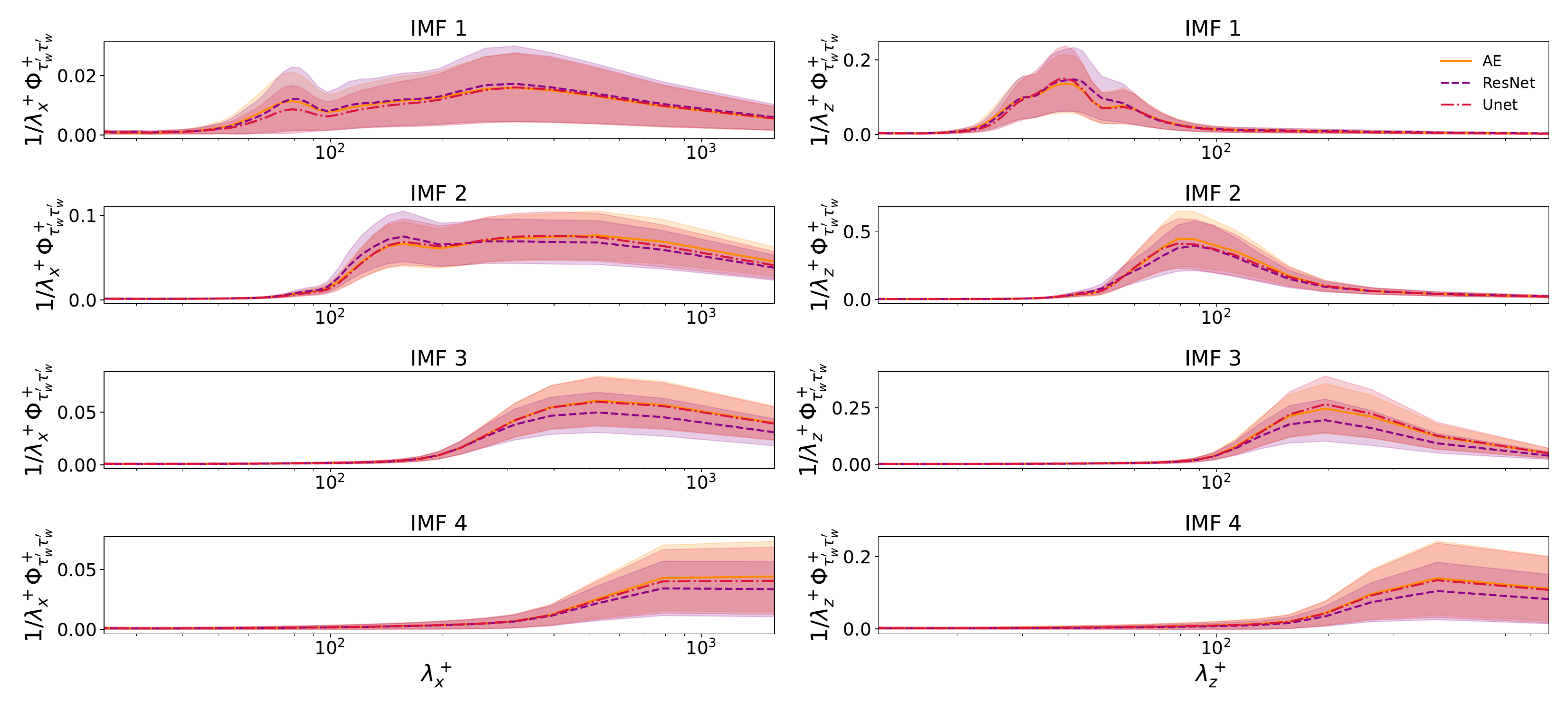}
  \caption{\textbf{Pre-multiplied power spectra of the IMFs of the experimental wall-shear stress predictions of the TCF at $\mathbf{Re_\tau \approx 1,000}$.} The left column provides spectra as a function of the streamwise wavelengths $\lambda_x^+$ and the right column as a function of the spanwise wavelengths $\lambda_z^+$. Lines indicate averages over all samples and the shaded regions are the corresponding standard deviations. Small differences can be observed between the three networks. For instance, the Unet deviates from the other distributions in IMF1, while the ResNet predicts a slightly lower intensity in IMF3 and 4.}
  \label{fig:spectra_EMD_exp_all}
\end{figure}

To investigate the networks' performance on experimental data in more detail, figure~\ref{fig:spectra_EMD_exp_all} provides the pre-multiplied power spectra of the predicted wall-shear stress fields. The trends of the distributions are similar, but in contrast to the numerical test data, on which all three performed equally well (figure~\ref{fig:spectra_EMD_DNS_all}), more distinct differences between the three architectures can be observed. The Unet's performance with respect to the smallest scales (IMF1) deviates from the two other networks, which explains why it was less accurate on an instantaneous level. For medium-size and large scales (IMF3 and 4), the ResNet based spectra indicate a slightly lower energetic level compared to the two other architectures. Thus, the ResNet presumably underestimates the intensity of large-scale flow structures in experimental settings.

\end{document}